\documentclass{emulateapj}

\newcommand{\kms}{\ensuremath{\,\mbox{km}\,\mbox{s}^{-1}}}

\newcommand{\Msun}{$M_{\odot}$}
\newcommand{\HI}{H{\sc i}\ }

\begin{document}
\title{The stellar population and interstellar medium in NGC 6822}
\author{W.J.G.~de~Blok}
\affil{Research School of Astronomy \& Astrophysics, ANU\\
Mount Stromlo Observatory, Cotter Road, Weston Creek, ACT 2611, Australia}
\email{edeblok@mso.anu.edu.au}
\and
\author{F. Walter}
\affil{Max Planck Institute f\"ur Astronomie\\
K\"onigstuhl 17, 69117 Heidelberg, Germany}
\email{walter@mpia.de}
 
\begin{abstract}
We present a comprehensive study of the stellar population and the
interstellar medium in NGC\,6822 using high-quality \HI data (obtained
with the Australia Telescope Compact Array) and optical
broad/narrow-band data (obtained with Subaru and the INT).  Our
H$\alpha$ observations are an order of magnitude deeper than previous
studies and reveal a complex filamentary network covering almost the
entire central disk of NGC\,6822. We find hitherto unknown H{\sc ii}
regions in the outskirts of NGC\,6822 and the companion galaxy. The
old and intermediate age stellar population can be traced out to radii
of over 0.6$^\circ$ ($> 5$ kpc), significantly more extended than the
HI disk. In sharp contrast, the distribution of the young, blue stars,
closely follows the distribution of the \HI disk and displays a highly
structured morphology.

We find evidence for an older stellar population in the companion
galaxy -- the current star formation activity, although likely to have
been triggered by the interaction with NGC\,6822, is not the first star
formation episode in this object. We show that the properties of the
giant kpc-sized hole in the outer \HI disk of NGC 6822 are consistent
with it being formed by the effects of stellar evolution.

\end{abstract}


\section{Introduction}
 
Dwarf irregular galaxies are ideal laboratories to study galaxy
evolution.  Their relatively simple structure, without dominant spiral
arms, bulges, and other complicating properties make it more
straight-forward to study the various physical processes related to star formation occuring in
these galaxies. Their low metallicities and gas-richness suggest that
they are still in an early stage of their conversion from gas into
stars.
 
NGC\,6822 is one of the closest gas-rich dwarfs in the Local Group.
Discovered by \citet{barnard}, it was studied in detail by
\cite{hubble} who showed that its distance placed it well outside our
Milky Way galaxy, making NGC 6822 the first object to be recognized as
an ``extra-galactic'' system (see also \citealt{perrine22}).
NGC\,6822 is not associated with any of the subgroups in the Local
Group \citep{mateo98, vandenbergh99} but is a member of an extended
cloud of irregulars known as the ``Local Group Cloud''
\citep{mateo98}.

Due to the small distance of $490\pm 40$ kpc \citep{mateo98} the
galaxy appears very extended on the sky: its optical angular diameter
is over a quarter of a degree; the H{\sc i} disk measures close to a
degree \citep{deblokwalter00}.   The galaxy has a total luminosity of $M_B\!=\!-15.8$
\citep{hodge91} and a
total H{\sc i} mass of $1.3 \times 10^8\ M_{\odot}$
(\citealt{deblokwalter00}; this work), making it relatively
gas-rich. It is a metal poor galaxy, with an ISM abundance of about
0.2 $Z_{\odot}$ \citep{evan89} and has a star formation rate of $\sim
0.06\ M_{\odot}{\rm yr}^{-1}$ (based on H$\alpha$ and FIR fluxes;
\citealt{mateo98,israel}). \citet{hodge80} found evidence for increased
star formation between 75 and 100 Myr ago, while \citet{gallart2}
showed that the star formation in NGC 6822 increased by a factor of 2
to 6 between 100 and 200 Myr ago. \citet{wyder}, from HST imaging,
found a spatially variable recent star formation rate in the central
parts of NGC 6822. This is all consistent with the mostly constant but
stochastic recent star formation histories often derived for dwarf and
LSB galaxies.  NGC 6822 can be regarded as a
rather average and quiescent dwarf irregular galaxy.

Recent optical studies have concentrated on the extended stellar disk
of NGC 6822. \citet{deblokwalter03}, \citet*{letarte03} and
\citet{komiyama03} all found that the stellar disk of NGC 6822 extends
much further out than the conventional optical appearance
suggests. Significant numbers of young stars were found throughout the
entire \HI disk.  Its inclination, gas-richness and proximity make it
an ideal candidate for studying the rotation curve, dark matter
content \citep*{weldrake03} and relation between the stellar
population and gaseous ISM. This is a major motivation for this study
of the stellar populations and neutral interstellar medium of
NGC\,6822.

In this paper we present the full \HI data set used earlier in
\citet{deblokwalter00} and \citet{deblokwalter03}.  We re-analyse the
Subaru optical data sets described in \citet{komiyama03}, reaching
significantly fainter limiting magnitudes. We discuss the distribution
of the old and young stellar populations, and find that NGC 6822 is
surrounded by a stellar halo measuring at least a degree in extent.
We discuss new deep, wide-field H$\alpha$ images of the entire \HI
disk, and find several faint hitherto unknown H{\sc ii} regions in the
outer parts of NGC 6822, including in the NW cloud and on the far rim
of the giant hole.  Finally, we study the stellar population in the
giant hole and find clear evidence for an age gradient, consistent
with the idea that star formation is capable of forming giant
kpc-sized holes in the ISM.  In a companion paper [de Blok \& Walter
  2005 (Paper II)] we use the data sets presented here to study in
detail the star formation threshold in NGC 6822.

In Sections 2, 3 and 4 we describe the various data sets used. Section
5 deals with the stellar populations in NGC 6822, and their relative
distributions. There we also discuss the properties of the NW cloud
as well as the
evolutionary history of the prominent super-giant shell.
In Sect.~6 we summarise our results.

\section{HI Observations}

\subsection{Single Dish}

NGC\,6822 was observed on 16 December 1998 using the ATNF Parkes
single-dish telescope in Australia. The narrow-band back-end system
\citep{haynes98} of the Multi-Beam system \citep{lss96} was used.
This set-up uses the central 7 beams.  A band width of 4 MHz with 1024
channels and a velocity spacing of 0.8 km\,s$^{-1}$ was used.
 
An area of $4 \times 4$ degrees was observed by scanning the telescope
over the sky at a rate of 1 degree per minute, outputting 14 spectra
(7 beams $\times$ 2 polarizations) every 5 seconds.  Frequency
switching was used to determine the bandpass correction.  The system
temperature was 24 K.
 
The individual bandpass-corrected single-beam, single-polarization
spectra were gridded into a data cube of 4.5 $\times$ 4.5 degrees
$\times$ 820 km\,s$^{-1}$ with a pixel size of 4$' \times 4' \times
1.6$ km\,s$^{-1}$. The effective beam size was 16.7$'$ (2.5\,kpc). The
total integration time per point was $\sim 35$ s per polarization,
resulting in a noise of 0.08 Jy beam$^{-1}$ per channel. The corresponding
column density sensitivity is $\sim 1.0 \cdot 10^{17}$ cm$^{-2}$ per
1.6 \kms\ channel for emission filling the beam. These numbers are
summarised in Table~\ref{tab:noises}.

With a systemic velocity of $-55$ \kms, the \HI emission from NGC
6822 partly overlaps with foreground Galactic emission. The velocity
range from $\sim -40 \kms$ to the upper end of the velocity cut-off of
the data cube used (at 50 \kms) was affected by this. In the range
$-14 \la V \la -3$ \kms\ ($\sim$ 6 channels) the Galactic emission was
strong enough to significantly hinder our ability to distinghuish the
NGC 6822 signal.  The strong Galactic emission profile had broad
low-level velocity wings, that showed itself as a slight enhancement
of the background (or rather, foreground) level of adjacent channels.

Ultimately, as we want to use this single-dish data to make a
zero-spacing correction to the interferometric data described in
Sect.~\ref{sec:atcadata}, we need to remove the extra Galactic
emission.  The data were corrected as follows: in all channels the NGC
6822 emission was blanked, and an annulus of data surrounding the
blanked emission was used to interpolate the foreground emission
across the blanked area.  These interpolated channel maps were
subtracted from the original channel maps, leaving only the NGC 6822
signal. The limited spatial resolution of the Parkes data meant that
the observed Galactic emission did not change on scales much smaller
than the extent of the NGC 6822 emission in each channel, thus
ensuring that no significant residuals were introduced by the fitting
and subtracting procedure.

The channels between $-14$ and $-3$ \kms\ were severely affected by
Galactic emission (the peak flux density of the Galactic emission was
generally two orders of magnitude larger than that of the NGC 6822
emission). Here we inferred the NGC 6822 emission from interpolated
values using a linear fit in the spectral direction to the
(foreground-corrected) emission in adjacent velocity channels.  
Figure \ref{fig:globprof} shows the uncorrected and corrected global
profiles.  These, and the total fluxes will be discussed in
Sect.~\ref{sec:globprof}.

\subsection{Interferometric data\label{sec:atcadata}}

NGC 6822 was observed in 1999 and 2000 with the Australia Telescope
Compact Array ({\sc atca}) in its 375, 750D, 1.5A, 6A and 6D
configurations. To cover the extent of the \HI disk of NGC 6822 as
derived from the single-dish data, we observed 8 pointings. These were
hexagonally packed along the major axis of NGC6822. This ensured that
most of the \HI disk of NGC 6822 was observed with a constant
sensitivity. In all cases we used PKS B1934--638 as the primary
calibrator. For the phase calibration we used PKSB 1954--388, which
was observed for 5 minutes every 40 minutes.  The central positions of
the pointings are given in Table~\ref{tab:atcapointings}.

The declination of NGC 6822 meant that in every observing slot the
total time on source available was approximately 10 hours.  We 
observed all pointings in mosaicking mode, taking care that the
$uv$ tracks were at least Nyquist-sampled. This meant that with the 375
array all 8 pointings could be observed in one observing session. With
the 750 array 4 pointings were mosaicked at once. Similarly, 2
pointings were mosaicked per slot with the 1.5 km array. Finally, with
the 6 km configurations all pointings were observed
individually. Details are given in Table~\ref{tab:atcaconfig}.  The
combination of all arrays gives a fairly homogeneous covering of the
$uv$ plane from the shortest baseline of 30 m to the longest 6 km
baseline, as Table~\ref{tab:atcaconfig} shows.  A correlator configuration
with a total bandwidth of 4 MHz and 1024 channels was used. This
yields a velocity spacing between channels of 0.8 \kms.  This
configuration is identical to that used for the single-dish data.

The data were checked for interference and calibrated using standard
procedures in the {\sc miriad} package. ``Dirty'' data cubes were
created using the ``mosaic'' option available in the {\sc miriad invert}
task. This option results in a super-uniform weighting scheme,
reducing side lobes in individual pointings prior to mosaicking.  

Our highest-resolution data cube was built up in several stages.  We
first constructed a low-resolution cube (with a resolution equivalent
to that which would have been obtained with the 375m array alone), and
in this cube identified {\sc clean} regions by hand, after one cycle
of 2$\sigma$ clipping.  The cube was then {\sc clean}-ed with the {\sc
  miriad mossdi} task to a level of about $0.5\sigma$ using a clean
beam of $396.4'' \times 96.0''$. In the following, we will indicate
the various resolutions by their minor-axis beam sizes. This lowest
resolution cube is thus referred to as `B96'.

The {\sc clean}-ed B96 cube was in turn clipped at $2\sigma$, and
checked for spurious noise peaks. The remaining regions were used as
{\sc clean} regions for the next higher resolution cube, B48. Again,
{\sc mossdi} was used for this. The resulting beam size was $174.7''
\times 48.0''$. The same procedure was then used to produce
respectively the B24, and finally B12 cubes.  We experimented with
various velocity resolutions, and found that 1.6 \kms\ gave the
optimum balance between resolution and signal-to-noise.  Clean beam
sizes, and noise levels and column density sensitivities for the 1.6
\kms\ cubes (which will be used in all further analysis) are
given in Table~\ref{tab:noises}.

\subsection{Global \HI profile\label{sec:globprof}}

Deriving the global \HI profile of such an extended object as NGC 6822
using interferometric data is not entirely
straight-forward. \citet{jorsater} pointed out that the usual
procedure of measuring fluxes in a cleaned map and assuming that the
clean beam applies to the whole map is not correct.  Rather, a map
consists of the sum of the restored clean components [with units Jy
(clean beam)$^{-1}$] and the residual map [with units Jy (dirty
beam)$^{-1}$].  If the map is not cleaned extremely deeply, or the
dirty beam is non-gaussian, discrepancies between the true flux and
the inferred flux can occur: the dirty beam usually has larger wings
than the clean beam, leading to an overestimate of the flux.  This is
especially relevant for extended objects.

Following \citet{jorsater} we thus compute the true flux
\[ G = {{D \times C}\over{D-R}}\]
where $G$ is the true flux, $D$ is the flux measured in the dirty
(uncleaned) maps, $C$ is the flux measured in the restored clean
components map, and $R$ is the flux measured in the residual map.  We
computed these fluxes for every channel map at each of the four
spatial resolutions, using the 1.6 \kms\ cubes.  Without this
procedure, fluxes in individual channel maps can differ by a
factor of two or more from the true flux.

\begin{figure}[t]
\epsscale{0.8}
\plotone{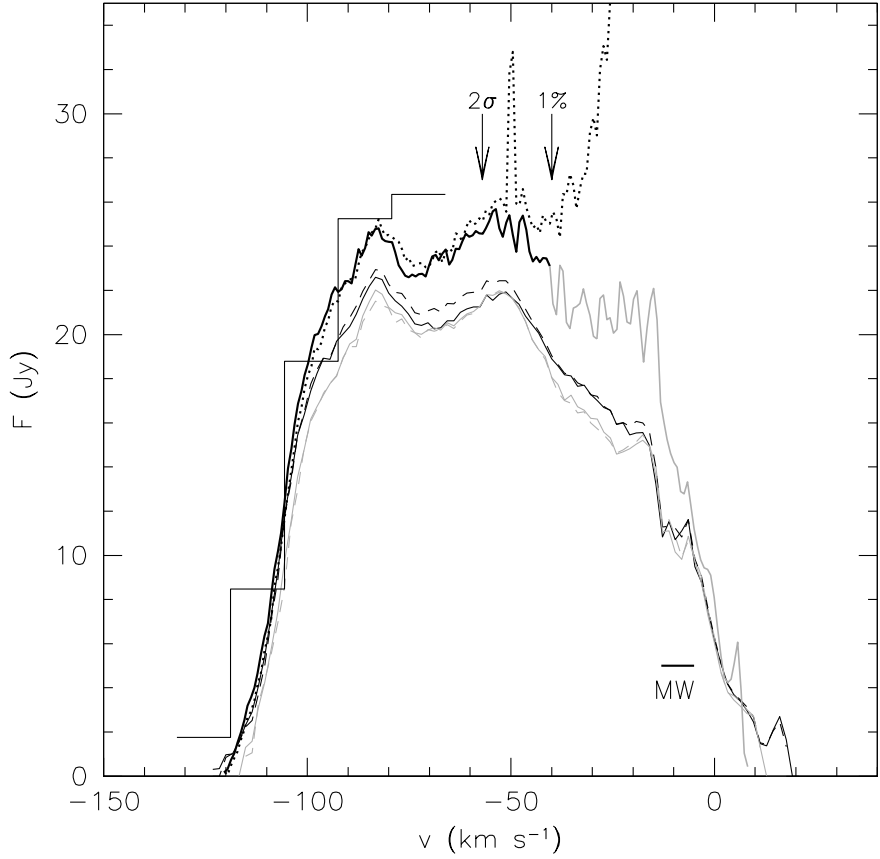}
\figcaption{NGC6822 global \HI profiles. The thick dotted line is the single-dish
profile not corrected for Galactic emission. The thick drawn line is
the same profile correced for Galactic emission. The black part of the
profile indicates the velocity range unaffected by Galactic foreground
emission, the grey part indicates the velocity range where the
corrections are significant.  The horizontal line marked `MW'
indicates the velocity range where in the single-dish data the signal
from NGC 6822 could not be distinghuished from Galactic emission. The
single-dish profiles in this velocity range were interpolated (see
text). The thin curves are the global profiles as derived from {\sc
atca} data. The black full curve represents the B96 data, black dashed
is the B48 data; the grey full curve is the B24 data; the grey dashed
curve represents the B12 data. The histogram in the velocity range
$-130\
\kms < V < -60\ \kms$ is the unaffected part of the HIPASS global
profile with a velocity resolution of 13.2 \kms. Arrows in the plot
indicate the velocities where the Galactic emission has fallen to 1
per cent of its peak value of $\sim 50$ Jy beam$^{-1}$, and where the
level of Galactic emission equals twice the noise in the single-dish
data cube.\label{fig:globprof}}
\end{figure}

Figure \ref{fig:globprof} shows the various global \HI profiles. The
four global profiles derived from the B96, B48, B24 and B12 data agree
very well with each other, with the expected trend that the B24 and
B12 data contain slightly less flux than the B96 and B48 data, as they
are less sensitive to extended low column density structures.

Also shown are the profiles as derived from the single-dish data.  We
plot the profile corrected for Galactic emission as well as the
original profile. A comparison shows that the uncorrected signal is
dominated by Galactic emission from $\sim -50 \kms$ and upwards.

The peak seen at $\sim -50 \kms$ in the uncorrected profile is
probably not caused by instrumental effects. The emission associated
with this peak fills a large fraction of the observed field, shows
structure, occurs in multiple channels and is seen to move across the
field with velocity.  It presumably corresponds with a Galactic \HI\
layer with a very low column density of $\sim 4.5 \cdot 10^{17}$
cm$^{-2}$ and a FWHM velocity width of $\sim 2 \kms$ (and therefore
resolved in the 0.8 \kms\ resolution single dish data).

The NGC 6822 single-dish flux is slightly higher than the
interferometric flux, indicating that emission is present at scales
larger than that detectable by the shortest baseline ($\sim 20$ m,
corresponding to 36$'$: although the shortest physical baseline in
Table 2 is 31 m, mosaicking decreases the effective shortest baseline
to $\sim 20$ m, see \citealt{ekers}). Incidentally, this makes the
interferometric observations excellent filters, that remove
most of the Galactic foreground emission, and elaborate procedures to
remove this emission from the {\sc atca} observations are not
necessary.

The higher flux in the single-dish profiles is not a result of
residual Galactic emission.  In Fig.~\ref{fig:globprof} we indicate
the velocity at which the Galactic emission has dropped to $1$ per
cent of its peak value ($\sim 50$ Jy beam$^{-1}$), as well as the approximate
velocity where the level of the Galactic emission equals 2 times the
noise in the unaffected parts of the single dish cube.  It is clear
that the differences in flux are already present in the unaffected
parts of the profile. The flux levels also agree well with those
derived from HIPASS data (also plotted in Fig.~\ref{fig:globprof} for
the unaffected part of the spectrum). Note that the HIPASS fluxes
shown here are uncertain by 10 per cent or so. This is due to the
HIPASS beam size changing slightly with source size and strength. In
Fig.~\ref{fig:globprof} we have adopted the case of a strong and very
($> 60'$) extended source. See Tables 2 and 3 in \citet{barnes01}
for more details.  Despite these uncertainties the single-dish fluxes
agree well with each other.

Table 3 contains the fluxes and \HI masses derived for the various
data sets.  We can put the following constraints on the \HI mass of
NGC 6822.  The interferometric profiles serve as a firm lower limit on
the \HI mass, independent of Galactic foreground emission, as these
observations have had by their nature most large-scale emission
filtered out. The average of the B96 and B48 masses is $1.20 \cdot
10^8$ \Msun.  The single-dish profile gives an \HI\ mass of $1.34
\cdot 10^8$ \Msun.

For $V < -40 \kms$, the \emph{shape} of the corrected single-dish
profile closely resembles that of the interferometric profiles.
Scaling the interferometric profile over this velocity range by a
factor of 1.12 makes the two profiles match. We assume that this
scaling corrects for missing flux not picked up by the interferometric
observations, and therefore adopt the single dish total \HI mass of
$1.34 \cdot 10^8$ \Msun as the \HI mass of NGC 6822 in the rest of
this paper.

\subsection{Zero-spacing correction and channel maps}

The {\sc miriad} task {\sc immerge} was used to combine the
single-dish and interferometric data, and apply the so-called
zero-spacing correction.

Figure~\ref{fig:chans} shows a selection of channel maps of the
zero-spacing corrected B12 cube with a velocity spacing of 1.6
\kms. Due to space limitations only every second channel is shown. The
velocity of each channel is shown in the top-left of each
sub-panel. The beam is shown in the bottom-right of the first
channel. Small residuals of Galactic emission are visible in the
channels in the range $-11 \la V \la -5$ \kms.

Primary beam corrections for each pointing are done implicitly during
the merging of the mosaic pointings and construction of the data
cubes. To give an idea of the area of the mosaic where the sensitivity
is highest and most uniform we show in contours the regions where the
sensitivity has dropped by to 90, 80 and 60 per cent of the central,
highest sensitivity. Comparison with the channel maps shows that
essentially all emission lies within the 80 per cent contour, well
within the equivalent of what in a single pointed observation would be
called the primary beam.

Several distinct features are visible in the channel maps. Firstly,
note the clear physical separation between the NW cloud and the main
body of the galaxy around $V \sim -100 \kms$. Within the NW cloud, the
high column-density \HI seems compressed into a ridge-like structure
($V\sim -104 \kms$), and there is a suggestion of hole-like structures
to the east of the ridge. This would fit in the picture of the NW
cloud as a separate star-forming entity as pointed out by
\citet{deblokwalter03}, \citet{letarte03} and
\citet{komiyama03}. Also note the low column-density plume that seems
to surround the NW cloud and is visible over the range $-113 \la V \la
-75$ \kms.

Moving up in velocity we see the main body of the galaxy appear. Here
there are clear signs of hole-like structures in the ISM (see e.g.\
the channel maps at $V=-78.4$ \kms\ and $V=-88.0$ \kms). The typical
size of these holes is $\sim 0.3$ kpc. The high-column density \HI\
again is again concentrated in ridge-like structures.

Past the systemic velocity of $-55$ \kms\ we start seeing signs of the super-giant shell
in the south-east part of the galaxy. The column density contrast in
this part of the galaxy is smaller than in the adjacent central part.
This is also the velocity range where residuals of Galactic emission
are visible, especially in the $-11.2$ and $-8.0$ \kms\ channels.

\subsection{Moment maps\label{sec:mommaps}}

\begin{figure*}[t!]
\plotone{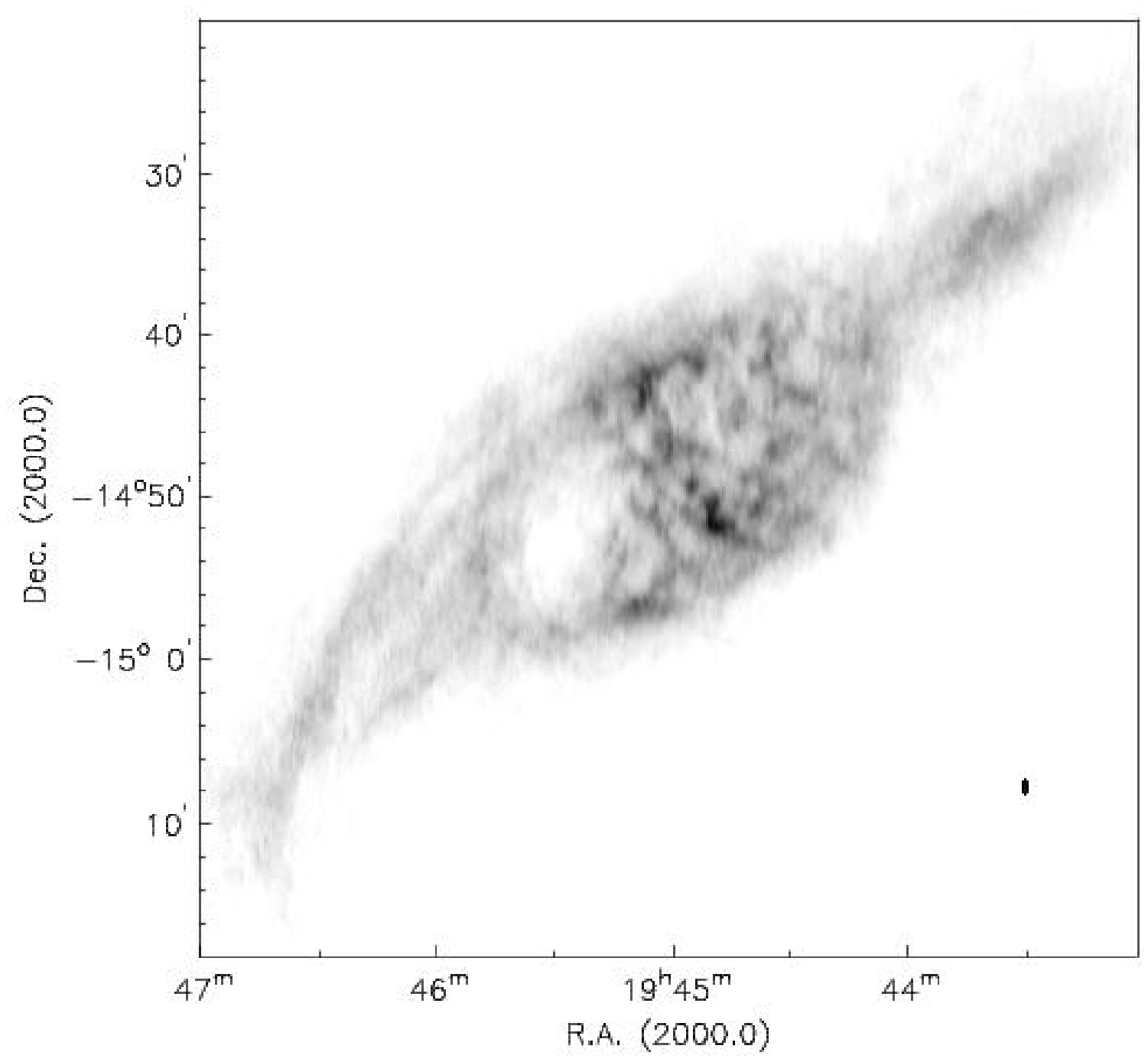}
\figcaption{Integrated \HI column density or zeroth moment map. The grayscale 
levels run from $1\cdot 10^{20}$ cm$^{-2}$ (white) to $4.3\cdot 10^{21}$
cm$^{-2}$ (black), which is also the maximum column density occuring
in this map. The beam of $42.4'' \times 12.0''$ is indicated in the
bottom-right corner. See \citet{weldrake03} for additional representations of this column density map.\label{fig:mom0}}
\end{figure*}

Figure~\ref{fig:mom0} shows the total \HI column density (zeroth
moment) map.  This map, as well as the velocity field, has also been
published in \citet{weldrake03}, but is reproduced here because of
the heavy use that will be made of them later in this paper.

The integrated \HI column density map was constructed by adding clipped
B12 channel maps. A similar method was used as when defining {\sc
  clean} regions.  The B96 cube was clipped at $2.5\sigma$, and the
remaining regions were used as a mask for the next higher resolution
cube, B48. The resulting masked B48 cube, was then clipped at
2.5$\sigma$, and in turn used as a mask for the B24 cube. The same
procedure was then used to produce the B12 cubes.

The resulting moment maps still contained some imperfections.  As a
second step we then isolated the high signal-to-noise ratio regions of
the zeroth moment map as follows. For uniformly tapered maps in
velocity $\sigma_{\rm tot} = \sqrt{N} \sigma_{\rm chan}$, where
$\sigma_{\rm tot}$ is the noise in a pixel in the integrated column
density map, $N$ is the number of channels contributing to that pixel
and $\sigma_{\rm chan}$ is the noise in one channel at that pixel. We
constructed a noise map corresponding with the column density map, and
used these two to isolate those pixels in the column density maps
where the signal-to-noise ratio was $>$ 4. These high signal-to-noise
ratio maps were used as masks for the higher-order moment maps.

The 4$\sigma$ limit in the total column density map is $\sim 1\cdot
10^{20}$ cm$^{-2}$. This is consistent with the noise limit \emph{per
  channel map} given in Table~\ref{tab:noises}, when one considers
that at the outer edge of the disk approximately 3 channel maps
contribute to one moment map pixel.

The same filamentary structure that was already visible in the channel
maps is also seen in the total \HI column density map. Signatures of
small shells are less clearly visible, mostly due to the integration
along the line of sight that tends to smear out structures due to gas
at different velocities.

This is an opportune moment to spend a few words on the definition of
the edge of the \HI disk of NGC 6822.  In the rest of the paper we
will often compare the extent of the \HI disk with that of other mass
components. It is therefore necessary to check whether it makes sense
to speak of the ``edge'' of the \HI disk. If any edge is simply caused
by observational sensitivity effects, where we lose the \HI column
density in the noise, then a sensible physical interpretation is not
really possible. If however we see a sharp decline in the \HI column
density well above the sensitivity limit of the data, then defining an
edge to the \HI disk is useful.

\begin{figure*}[t!]
\plotone{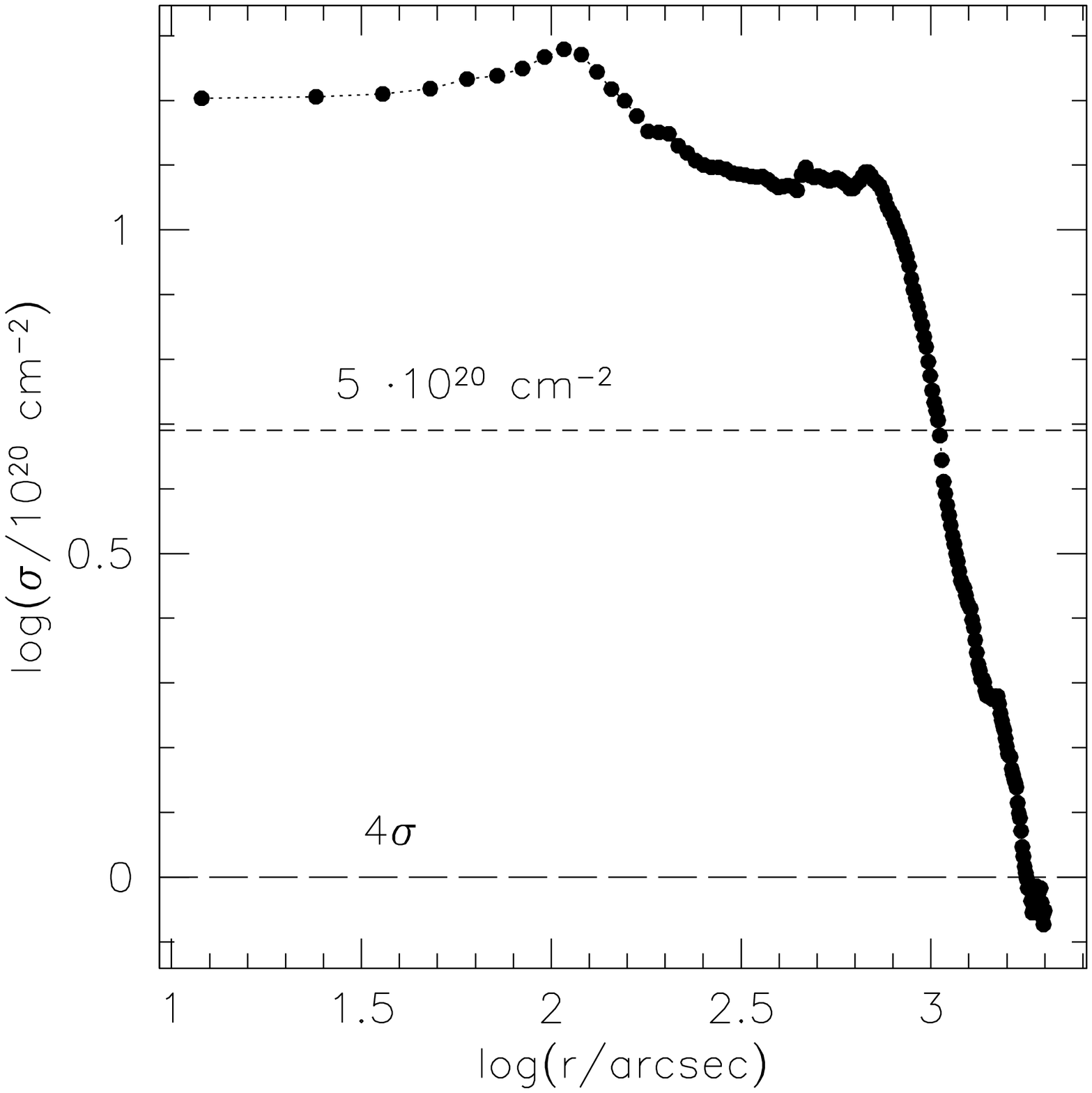} \figcaption{Azimuthally averaged \HI profile. The
  $4\sigma$ level in a single channel map is indicated as well as the
  column value of $5 \cdot 10^{20}$ cm$^{-2}$ (not corrected for
  inclination) that we use to indicate the edge of the \HI disk. Note
  that this occurs well above the sensitivity limit of the data.
\label{fig:hiedge}}
\end{figure*}

As stated above, the $4\sigma$ column density limit of the \HI column
density map is $1\cdot 10^{20}$ cm$^{-2}$.
Figure~\ref{fig:hiedge}\ shows a plot of the observed azimuthally
averaged \HI column density (derived using the tilted-ring parameters
given in \citealt{weldrake03}, also see below) with the $4\sigma$
column density limit indicated.  We see that over the first $\sim
800''$ ($\log R < 2.9$) (corresponding with the central \HI disk) the
column density only drops by a factor of $\sim 1.4$. Beyond that, over
a similar radial range, the column density drop by over a factor of
10.  The steep drop at the outer radii can be said to define some sort
of edge, a sharp transition from the high-column density regime
defining the shape of the galaxy to a more diffuse, possibly more
pervasive state.  In the following we will use a column density level
of $5 \cdot 10^{20}$ cm$^{-2}$ to indicate the edge of the \HI disk.

\begin{figure*}[t!]
\plotone{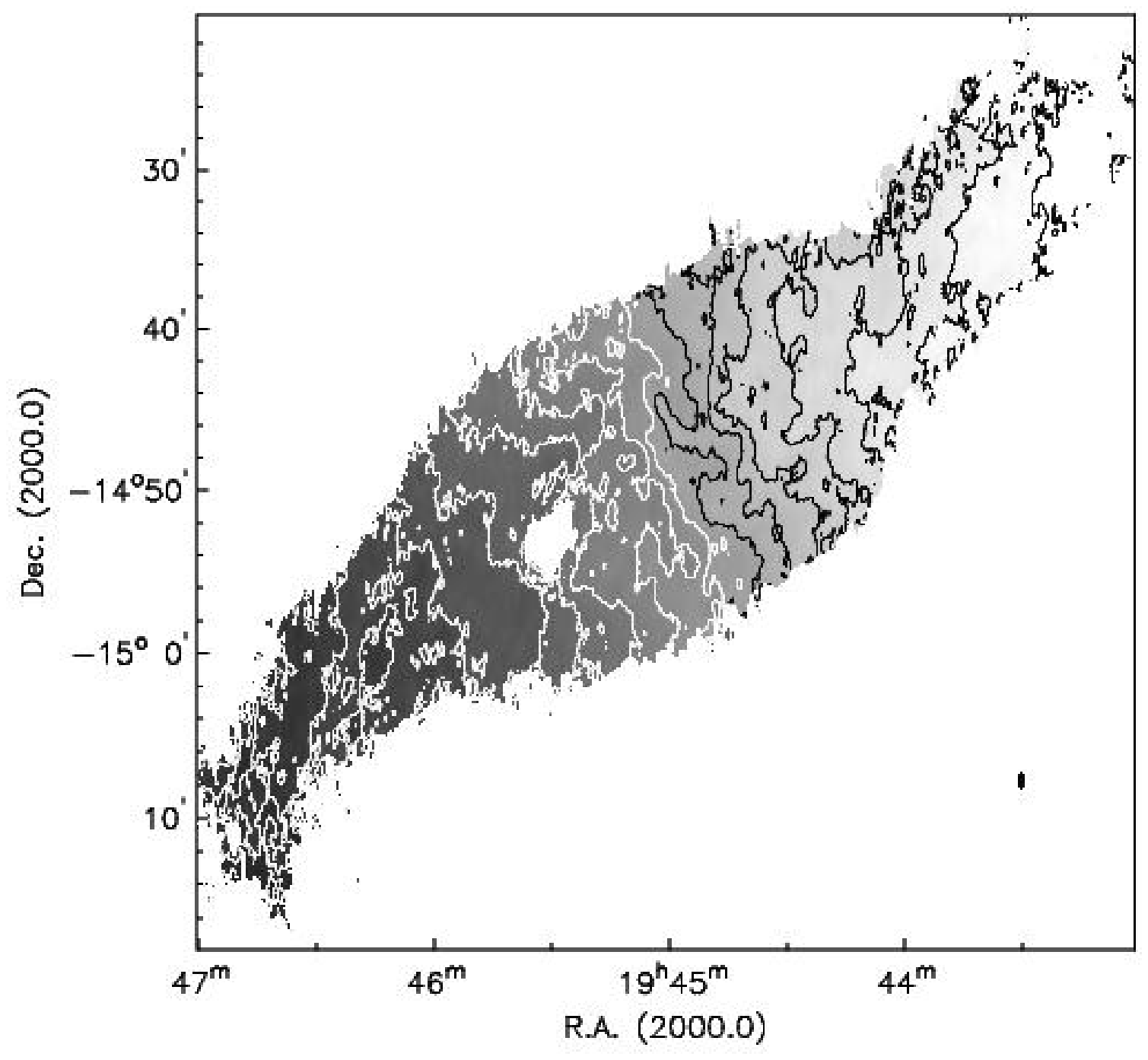}
\figcaption{Velocity field or first moment map. The black contours
run from a central value of $-58$ \kms\ to $-100$ \kms\ in the NW,
decreasing in steps of 7 \kms. The white contours run from a central
value of $-51$ \kms\ to a value of $+12$ \kms\ in the SE, increasing
in steps of 7 \kms.\label{fig:mom1}}
\end{figure*}

Figure~\ref{fig:mom1} shows the corresponding velocity field (first
moment map).  NGC 6822 is in fairly regular solid-body rotation. A
full analysis of the kinematics of NGC 6822, and an analysis of its
rotation curve has been presented in \citet{weldrake03}. They find
that NGC 6822 has a slowly rising, almost solid body like inner
rotation curve that flattens towards larger radii. They also find NGC
6822 is dark-matter dominated, with a ratio of visible (baryonic) to
dark matter of $\sim 0.09$ within the last measured point of the
rotation curve ($\sim 5$ kpc).  We refer to \citet{weldrake03} for a
more extensive discussion on the properties of the dark matter halo of
NGC 6822.  We will adopt the orientation parameters derived in that
paper.

\section{H$\alpha$ observations}

\subsection{The data}
NGC 6822 was observed with the 2.5m INT telescope at La Palma during
12-15 July 2004. We used the Wide Field Camera with a pixel scale of
$0.33$ arcsec pixel$^{-1}$. Three pointings were observed covering the
entire extent of the \HI disc.  Each pointing was observed for 4800
seconds. Short $R$-band exposures of 300s per pointing were also
obtained to determine the continuum. 

Figure~\ref{fig:halfa} compares the observed H$\alpha$ emission with
the extent of the \HI disc.  Obvious defects due to the subtraction of
the continuum have been removed. The image has been median-filtered
with a $9\times9$ pixels window to enhance the extended, low surface
brightness structures, and depress the continuum-subtraction
residuals. Due to the variable observing conditions an absolute flux
calibration could not be obtained. We instead used the list of NGC
6822 H{\sc ii} regions by \citet*{hodge89} to obtain a calibration.
Our observations are deeper than the ones presented in
\citet*{hodge88}, \citet{hodge89} and \citet{chyzy03} and cover a
larger area.  The limiting H$\alpha$ flux surface density of the
\citet{hodge89} observations is given as $2 \cdot 10^{-17}$ ergs
cm$^{-2}$ s$^{-1}$ arcsec$^{-1}$. The limiting flux surface density in
our observations is $\sim (1.1 \pm 0.2) \cdot 10^{-18}$ erg cm$^{-2}$
s$^{-1}$ arcsec$^{-1}$, where the uncertainty is mainly due to the
bootstrapping used to put our fluxes on the \citet{hodge89} scale.
Our data thus go a factor 10 deeper than previous observations.

\citet{hodge89} give a value of $1.8 \cdot 10^{39}$ erg s$^{-1}$ for
the total H$\alpha$ flux of NGC 6822\footnote{\citet{hodge89} use
  different values for the distance modulus and foreground
  extinction. We have converted the values given in that paper using
  our adopted \citet{gallart1} values.}.  The value we derive from our
observations is $(2.0 \pm 0.4) \cdot 10^{39}$ erg s$^{-1}$, after a 5
per cent correction for [N{\sc ii}] emission.  The total H$\alpha$
flux is thus slightly larger than the \citet{hodge89} value,
consistent with the fact that we have surveyed a larger area to a
larger depth.

\subsection{H$\alpha$ morphology} 

The H$\alpha$ emission is found throughout almost the entire main \HI
disc to the west of the hole. It is distributed in a filamentary
network that is clearly shaped by the effects of shocks and winds.
Many of the fainter H{\sc ii} regions catalogued in
\citet{hodge88,hodge89} as separate regions are in fact part of
continuous larger loops and filaments, as are the bright
H{\sc ii} regions.

Our H$\alpha$ data cover the outer \HI disk, a part of NGC 6822 not
previously observed to this depth.  The outer
regions do in fact contain a small number of low-luminosity H$\alpha$
regions, showing that star formation is proceeding even in the outer
parts of the galaxy.  Outside the area surveyed by
\citet{hodge88,hodge89} we find a small number of new H{\sc ii} regions that are
not obviously connected with the central H$\alpha$ network. The most
note-worthy of these are briefly described here.

\begin{figure*}[t!] 
\plotone{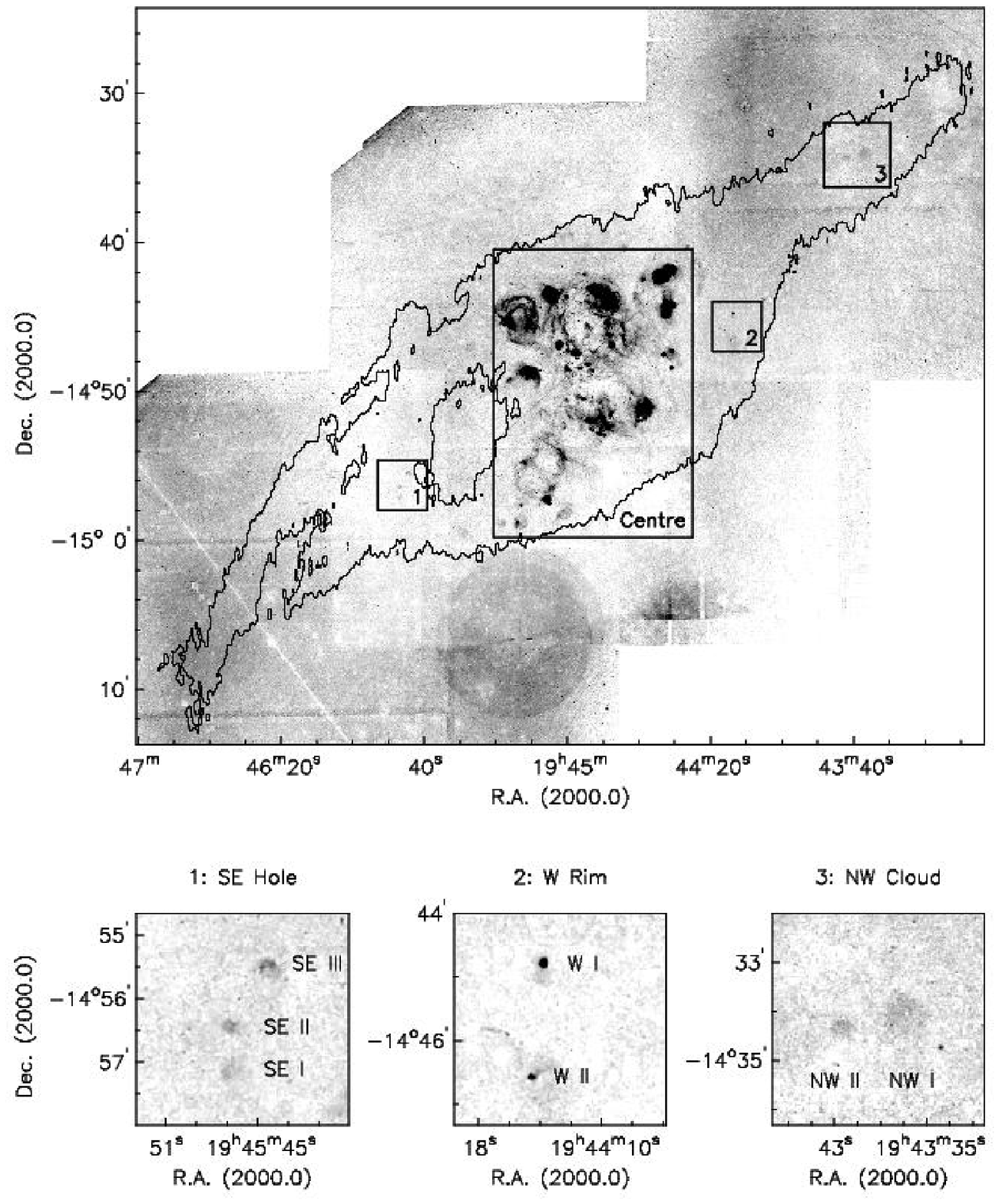} \figcaption{The H$\alpha$ distribution in NGC6822.
  The top panel shows the entire extent of our H$\alpha$ data.  The
  contour indicates the edge of the \HI disk at $5 \cdot 10^{20}$
  cm$^{-2}$ (not corrected for inclination). The large box labeled
  ``Centre'' marks the central area shown in more detail in
  Fig.~\ref{fig:halfa_cen}.  The boxes labeled ``1'', ``2'' and ``3''
  show the outlying areas where new HII regions were found. These
  areas are shown in more details in the small panels at the
  bottom.\label{fig:halfa}} \end{figure*}

\emph{NW cloud:} we find two new H{\sc ii} regions in the NW cloud.  These
have fluxes of respectively $\log(F_{H\alpha}) = 4.7$ (NW {\sc i}) and
$\log(F_{H\alpha}) = 4.3$ (NW {\sc ii}), where the fluxes are in units of
$10^{-18}$ ergs cm$^{-2}$ s$^{-1}$, following the notation given in
\citet{hodge89}. See Fig.~\ref{fig:halfa} for identifications.  These
fluxes are comparable to those of the lowest luminosity regions
catalogued by \citet{hodge89}, but the new regions tend to have a
somewhat lower surface brightness.

\emph{SE Hole:} we find three new H{\sc ii} regions on the opposite
(eastern) rim of the large hole in the \HI distribution. Their
respective fluxes are $\log(F_{H\alpha}) = 4.0$ (SE {\sc i}), $4.0$
(SE {\sc ii}), and $4.2$ (SE {\sc iii}). See Fig.~\ref{fig:halfa} for
identification.  These regions are also among the very faintest in NGC
6822.

\emph{Western Rim}: to the west of the main H$\alpha$ complexes we
find two solitary, compact H{\sc ii} regions. Their fluxes are
$\log(F_{H\alpha}) = 4.7$ (W {\sc i}) and $4.0$ (W {\sc ii}),
respectively. Again, see Fig.~\ref{fig:halfa} for identification.

These new regions do not increase the total H$\alpha$ flux emitted by
NGC 6822 significantly, but they do show that low-level star formation
does occur in the outer disk of NGC 6822. If the current state of NGC
6822 is typical, then they do strongly suggest that the outer stellar
disk of NGC 6822 (as described below and in \citealt{deblokwalter03},
\citealt{letarte03} and \citealt{komiyama03}) was built up \emph{in
  situ}, a process that is still happening today.

\section{Optical Broadband Data}

We used two sets of archival CCD data taken with the Subaru Prime
Focus Camera (Suprimecam) on the 8.2m Subaru Telescope on Mauna Kea,
Hawaii. Suprimecam consists of $5 \times 2$ CCDs of $2048 \times 4096$
pixels each, giving a total size of 10k $\times$ 8k pixels, or a 34$'$
by 27$'$ field of view with a 0.2$''$ pixel size \citep{miyazaki02}
 
The first data set is described in detail in \citet{komiyama03}, and
consists of deep $B$, $R_C$ and $I$ images spanning the entire \HI
disk of NGC 6822.  The second set consists of shallow exposures using
the same filters of the central optical part of NGC 6822 taken in as
test observations in 2000 during the early days of Suprime-Cam.  Both
data sets are essential to this analysis. As described below, the
large light-gathering power of Subaru meant that stars down to quite
faint magnitudes were already saturated, and combination with a
shallow data set is essential to get a complete view of the stellar
population.  Both sets are described in more detail below.
 
\subsection{The deep data set}
 
\begin{figure*}[t!]
\plotone{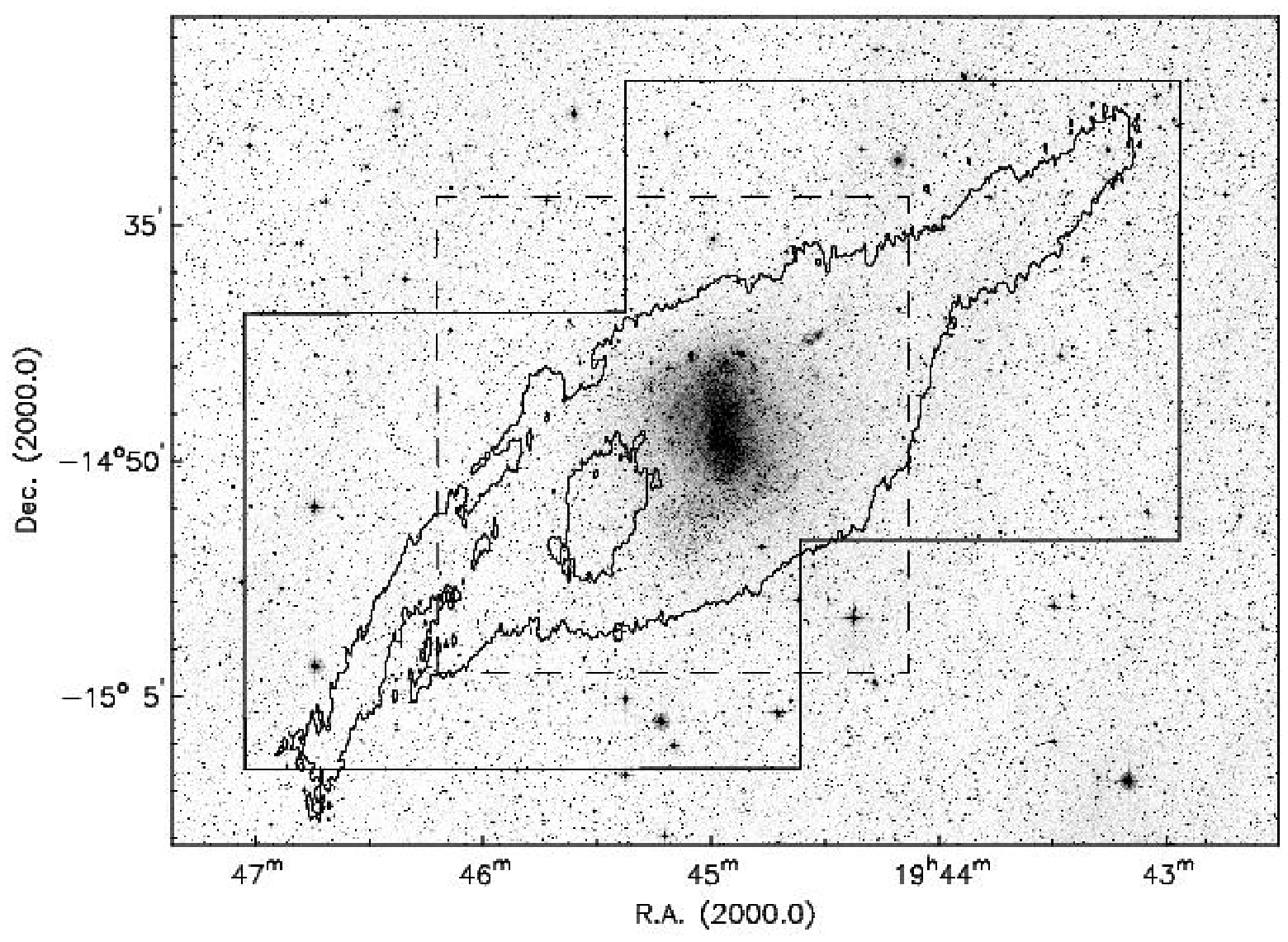}
\figcaption{The 
areas covered by both sets of Suprimecam observations overlaid on a
DSS image of the field around NGC 6822. The large full box shows the
field covered by the deep survey. The smaller dashed box shows the
field of the shallow survey. The merged catalog covers the overlap
between both sets. The contour indicates the edge of the \HI disk at $5
\cdot 10^{20}$ cm$^{-2}$ (not corrected for inclination).
\label{fig:subarufield}}
\end{figure*}

This data set consist of two pointings each in $B$, $R_C$ and $I$. The
first pointing covers the NW part of the disk as well as the NW cloud.
The second pointing covers the tidal tail and the SE part of the disk.
Together, as mentioned, they cover the whole of the \HI disk of NGC
6822 (see Fig.~\ref{fig:subarufield}).  The images were taken on 15
and 19 October 2001, and are described in \citet{komiyama03}, where a
first analysis of the $B$ and $R_C$ data is also presented. We used
the {\sc iraf mscred} package to perform the standard CCD reductions.
The total exposure times per pointing were 1440s in $B$, 2160s in $R$
and 960s in $I$. The USNO-B catalogue was used to determine the
coordinate system of these images.  The final RMS uncertainty in the
plate distortions as well as the final World Coordinate System was
$\sim 0.18"$.
 
The {\sc iraf} version of the {\sc daophot} package was used to find
all stellar objects in the images and produce lists of their
cooordinates, magnitudes and other parameters. We used modified
versions of the reduction and analysis scripts used by the Local Group
Survey group 
\citep{massey}. The average seeing in the three bands was 
$0.8''$ ($B$), $0.7''$ ($R$) and $0.6''$ ($I$), respectively.  
Standard stars were used to provide absolute calibration. The RMS in
the standard star solutions was typically $\sim 0.03$ mag.

All objects with peak values more than $3\sigma$ above the sky were
catalogued.  The separate $B$, $R$ and $I$ catalogues of the two
pointings were merged by cross-correlating and retaining only those
objects that were present in all three bands.  We assumed a match if
the coordinates agreed to within $0.5''$. For the very small number of
multiple matches within the error radius the brightest object was
always chosen.  Objects with a ``roundness'' (a {\sc{daophot}}
parameter that can be used to distinghuish stars from non-stellar or
blended objects) that differed by more than unity from the average
roundness of all objects in the field were rejected.  After this, the
NW field contained $326298$ objects in $B$, $589788$ in R and $643304$
in I. For the SE field the numbers are 363962 $(B)$, 644213 $(R)$,
552568 $(I)$. The cross-correlated catalog of both fields combined
with detections in all three bands contained 366162 objects.

\begin{figure*}[t!]
\epsscale{0.65}
\plotone{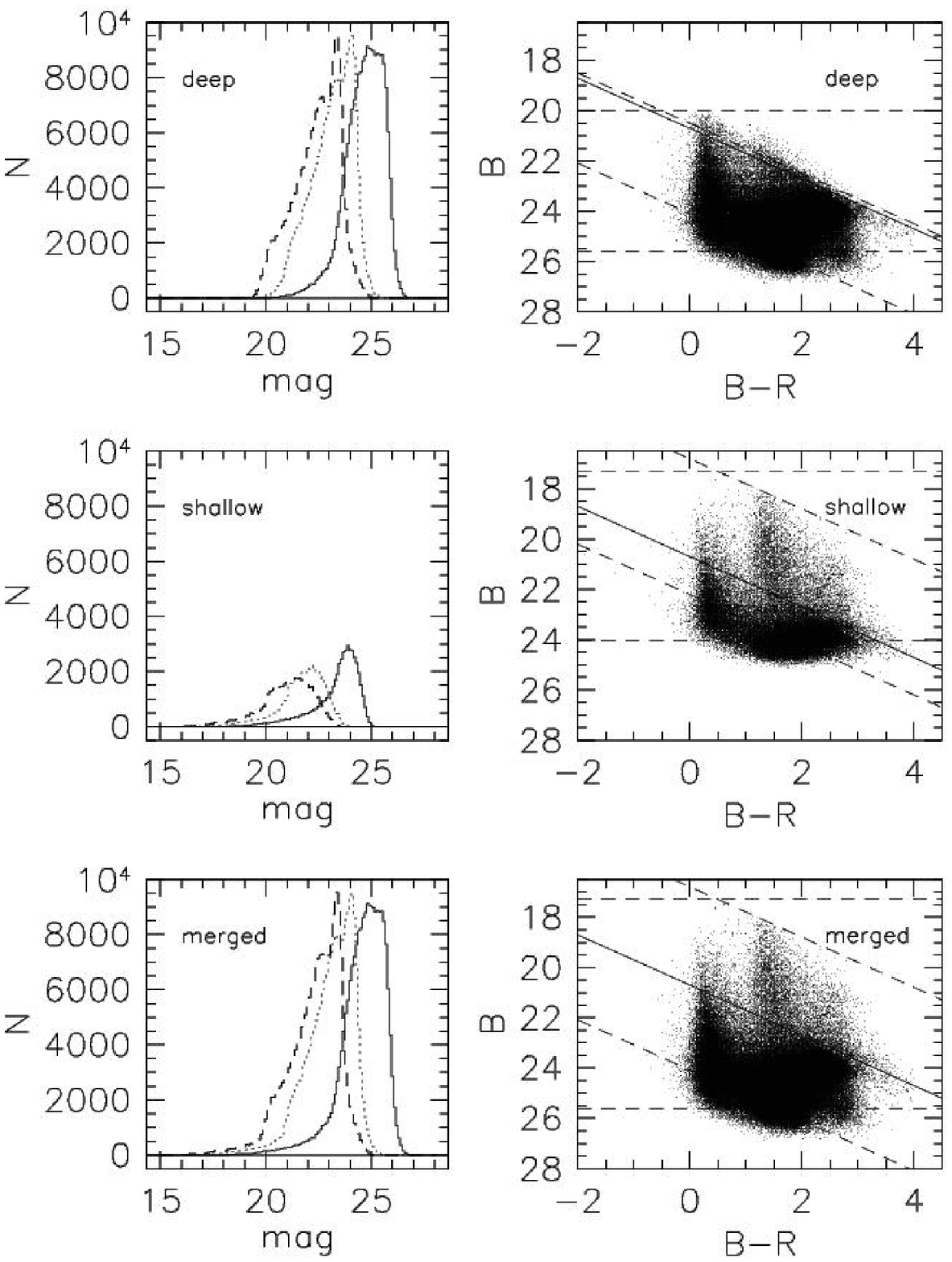}
\figcaption{Magnitude distributions and CMDs based on the two Subaru data sets.
\emph{Top row:} The deep catalogue. Note that we only show the data from the
overlap area with the shallow survey. \emph{Middle row:} The shallow
catalogue.  \emph{Bottom row:} The merged catalogue.  The \emph{left panels}
show the magnitude distributions in the three catalogues. The full
histograms shows the $B$-band data, the dotted histograms the $R$-band
and the dashed histograms the $I$-band data.  The \emph{right panels} show
the $B$ \emph{vs} $B-R$ CMDs for the three catalogues.  The horizontal
dashed lines show the completeness limits in $B$, the diagonal dashed
lines the limits in $R$. For the deep survey (top) these limits occur
at $m_B = (20.0,25.6)$ and $m_R = (20.5,24.1)$. The shallow survey
limits (middle) occur at $m_B = (17.3,24.1)$ and $m_R =
(16.8,22.2)$. The limits for the merged catalogue (bottom) are the
extremes of these limits. The full diagonal line at $m_R=20.7$ indicates
the magnitude cut applied to merge the shallow and deep catalogues. See
text for more details.
\label{fig:shallowdeep}}
\end{figure*}

In our analysis we will limit ourselves to only those ``high-quality''
objects with an uncertainty in the magnitude $< 0.2$ mag in all three
bands. To isolate true stellar objects we also insist that these
objects have an absolute value of the ``sharpness'' (one of the {\sc
  daophot} output parameters describing how ``stellar'' or peaked an
object is) less than unity.  These conditions then reduce the number
of objects in the final deep catalog to 251927.  The top row in
Fig.~\ref{fig:shallowdeep} show the histograms of apparent magnitudes
in the three bands, as well as a colour-magnitude diagram (CMD).  For
reasons of compatibility with the shallow survey (discussed below) we
only show the data from the overlap area with that survey.  The data
set becomes incomplete at the faint-end side at apparent magnitudes
$(m_B,m_R,m_I) = (25.6, 24.1, 23.4)$. These are the magnitudes where
the histograms peak and then rapidly drop towards fainter
magnitudes. At the bright end side the catalogue is limited by
saturation of the CCDs. This happens at magnitudes $(m_B,m_R,m_I) \sim
(20.0, 20.5, 19.5)$. These latter values are uncertain by $\sim 0.5$
mag, as they depend on bias and sky levels, which differ from chip to
chip and from observation to observation.

\subsection{The shallow data set}

This data set consists of a single pointing towards the optical centre
of NGC 6822. These are early observations (PI: S.~Miyazaki) taken in
June 2000 when Suprime-Cam only had 8 chips available. Some chips
suffered from large readout noise. Observations times were 600s in
$B$, 480s in $R$, and 330s in $I$. The seeing was $0.8''$ in $B$,
$0.8''$ in $R$ and $0.6''$ in $I$. These observations do not cover the
entire \HI disk, but are still useful as a complement to the deep data
set, as described below.  The area of the shallow survey is shown in
Fig.~\ref{fig:subarufield}.

Catalogues of stellar objects were constructed in an identical manner
as with the deep data set. The $B$-band catalogue initially contained
330830 objects, the $R$ catalogue 273078 objects and the $I$ catalogue
530826 objects. A merged list containing all objects with a detection
in all three bands contained 114179 objects.  The final catalog of
``high-quality'' stars as defined above contained 53008 objects.

No standard star observations were available for this run, so we used
the deep catalog to boot-strap the flux scale of the shallow survey.
The RMS uncertainties in this boot-strap are $\sim 0.15$ mag in $B$,
$\sim 0.14$ mag in $R$ and $\sim 0.13$ mag in $I$. This scatter is
mostly caused by the large read-out noise of some of the CCD chips
used to obtain the shallow data set. However, we will only use the
brightest stars from the shallow set, where the uncertainty in the
magnitudes is about a factor two lower than for the total shallow
survey.

The centre row of Fig.~\ref{fig:shallowdeep} shows the distribution of
magnitudes as well as the CMD.  For these shallow data incompleteness
at the faint end occurs at $(m_B,m_R,m_I) = (24.1, 22.2,
21.7)$. Saturation occurs at $(m_B,m_R,m_I)\sim (17.3, 16.8, 15.9)$.
These limits are again indicated in the figure.

\subsection{The merged data set}

\begin{figure*}[t!]
\plotone{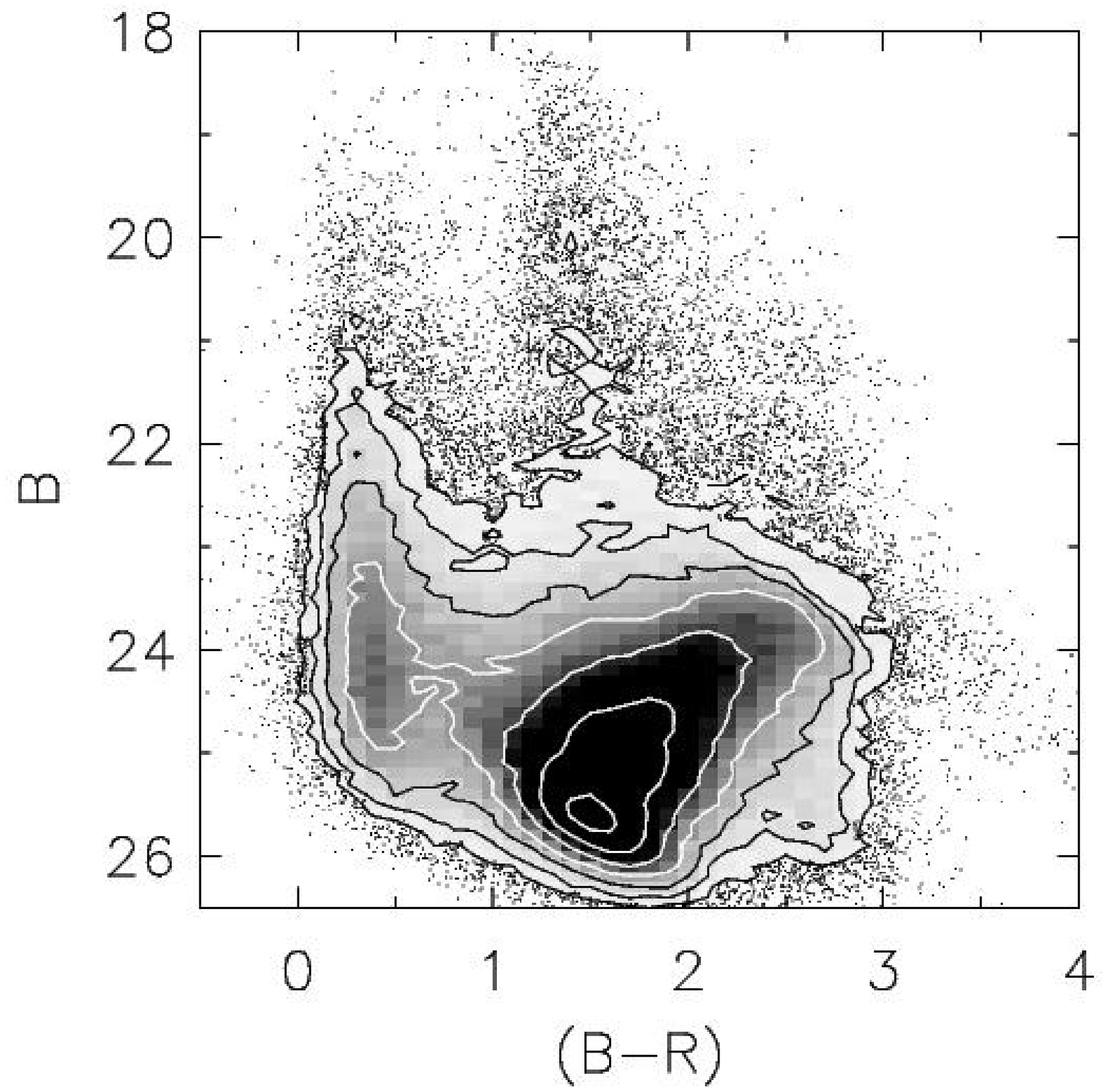}
\figcaption{Density contours of the CMD diagram of the merged
  catalogue.  The CMD was binned in intervals of  $0.1 \times
  0.1$ mag.  Contour levels are 20, 40, 80, 160, 320, 640, 1280 stars
  per bin.  For CMD surface densities less than 20 stars the
  individual stars are plotted.
\label{fig:surfdensCMD}}
\end{figure*}

In addition to the deep and the shallow catalogues we also produced a
merged catalogue containing all high-quality detections in both
catalogues, restricted to the overlap area between the data sets (see
Fig.~\ref{fig:subarufield}). Examination of the bright and faint
magnitude limits of the deep catalogue showed that the $R$ band data
are most restricted in dynamic range: stars saturate in the $R$ band
well before they saturate in $B$ or $I$. The best way to construct a
merged catalogue is thus to apply a magnitude cut in $R$.  We chose a
cut-off of $m_R = 20.7$, thus taking stars fainter than this from the
deep catalogue and brighter than this from the shallow catalogue.
This merged list contained 218438 objects.  The bottom row in
Fig.~\ref{fig:shallowdeep} shows the combined magnitude distributions
and CMDs. The line $m_R=20.7$ is also indicated. The surface
densities of stars in the CMD are continuous across the line,
indicating neither the deep nor the shallow catalogues are severely
incomplete with respect to each other at this magnitude.
Figure~\ref{fig:surfdensCMD} shows the surface density of stars
in the merged CMD which more clearly shows the structure in the
densely populated parts of the diagram.

\section{The stellar population of NGC 6822}

Though partly based on the same data as the analysis in
\citet{komiyama03} the CMD presented here has a larger dynamic range
and probes to fainter magnitudes. By adding in the shallow data we are
now also probing the O and B star regime that \citet{komiyama03} were
unable to probe.  This is the first time the stellar population of NGC
6822 has been mapped over such a large extent and depth.  We are
therefore in an excellent position to revisit and elaborate some of
the results from the papers of
\citet{komiyama03}, \citet{letarte03} and \citet{deblokwalter03}.

\subsection{The Stellar Extent of NGC 6822}
The first question one may ask is what is the total stellar extent of
NGC 6822?  Observations of \citet{letarte02} show the presence of
carbon stars outside the \HI disk, and as carbon stars generally trace
an intermediate age population this could indicate the presence of an
extended stellar component.

\begin{figure*}[t!]
\plotone{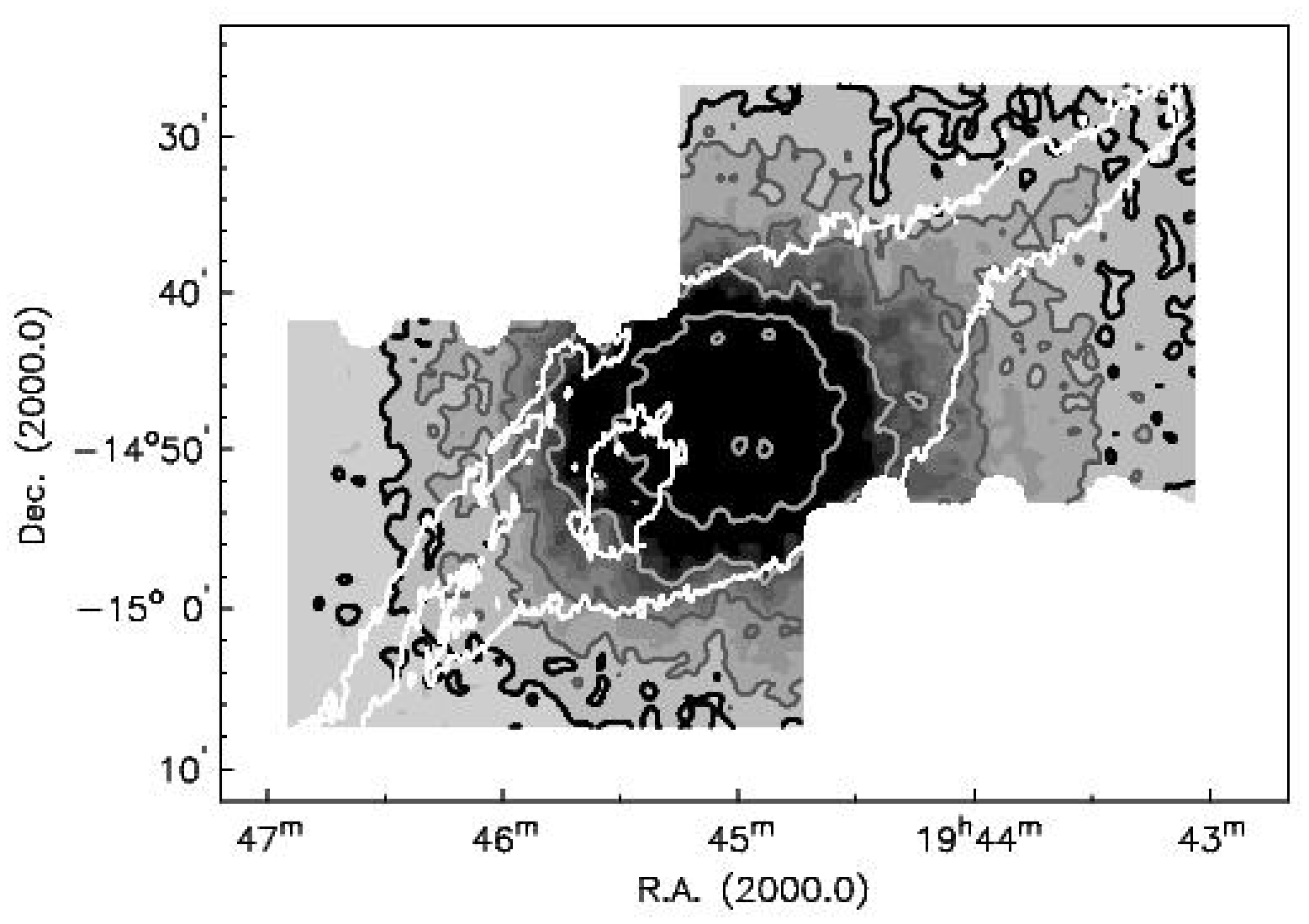}
\figcaption{Stellar surface density distribution based on the deep catalog.
Surface densities shown here were derived by counting stars in $12''
\times 12''$ boxes, smoothing the resulting distribution to  $48''$.
The thick black contour represents a smoothed surface density of
1.1. We take surface densities lower than this number to describe the
field environment. The grey contours show the 2, 4, 8, 16 and 32
surface density levels. 
The white contour indicates the edge of the \HI disk at $5
\cdot 10^{20}$ cm$^{-2}$ (not corrected for inclination).
Note that NGC 6822 is much
more extended to the NW as it is to the SE.
The white features along the edges of the field are artefacts of the
binning and smoothing procedures used.
\label{fig:surfdens}}
\end{figure*}

To determine the total extent of the stellar population of NGC 6822 we
have taken the deep catalog, and counted the total number of stars in
$12'' \times 12''$ boxes. Note that due to the saturation limits of
the deep catalog, a significant fraction of the Milky Way foreground
has effectively already been removed.  This surface density image was
then smoothed with a gaussian with FWHM $= 48''$.  This smoothed
stellar surface density image is shown in Fig.~\ref{fig:surfdens}. The
very high stellar density in the inner parts, as well as the omission
of the shallow catalogue, means that there will be some incompleteness
there, but in the outer, lower density parts this image should be a
fair representation of the total stellar surface densities there.  The
stellar surface density distribution is centered on NGC 6822,
showing that we are indeed looking at its stellar population, and not
Galactic foreground stars.  The stellar distribution
of NGC 6822 is much more extended than the \HI disk.

\begin{figure*}[t!]
\plotone{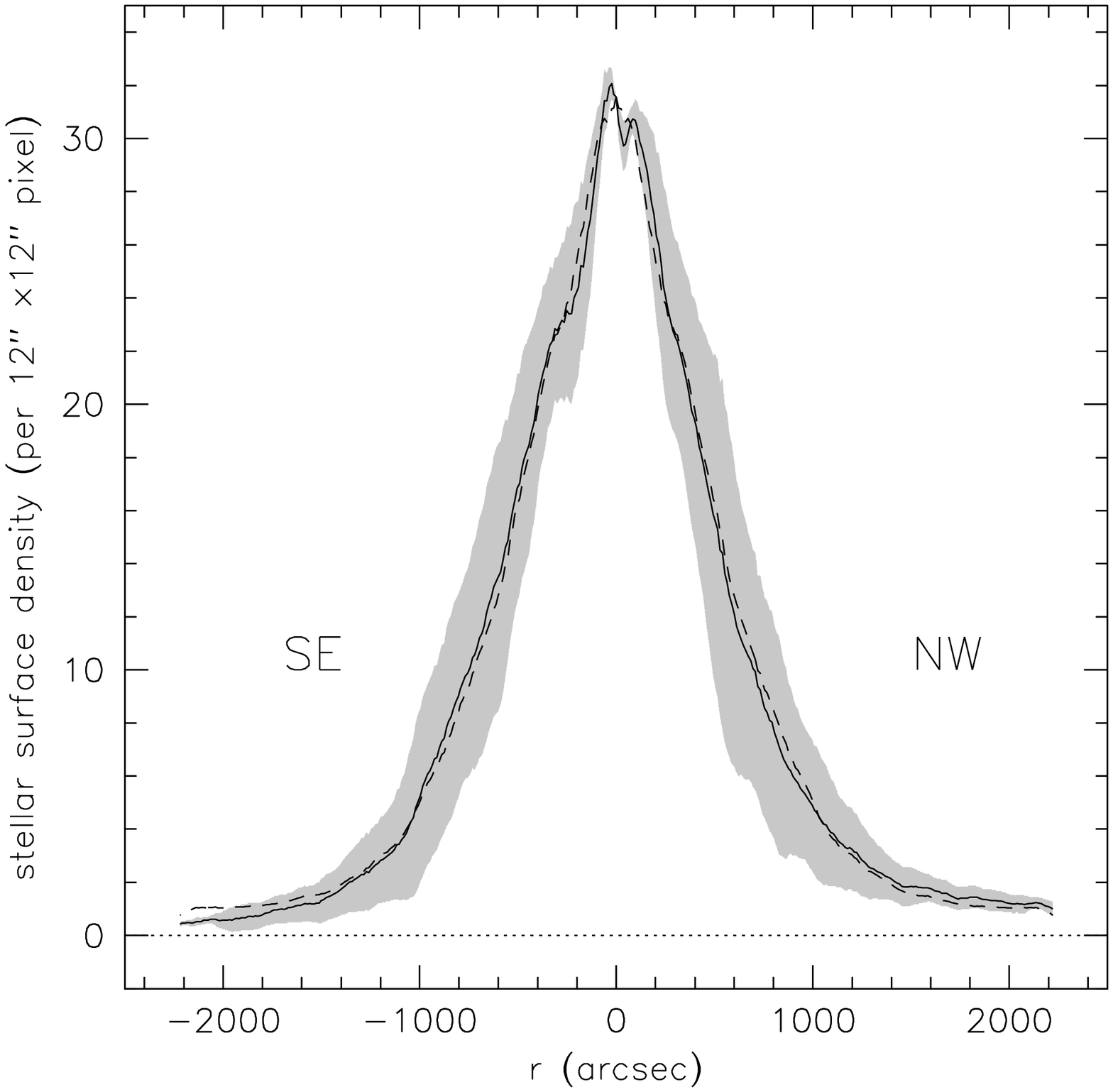}
\figcaption{Azimuthally averaged stellar surface density distribution as
derived from Fig.~\ref{fig:surfdens}. The tilted ring parameters as
described in Sec.~\ref{sec:mommaps} were used. The dashed line
represents the surface density averaged over both sides of the
galaxy. The full lines represent the profiles of the NW
(approaching) and SE (receding) side, as indicated in the plot.  The
grey areas indicate the 1$\sigma$ variation in the stellar surface
density in each annulus.  \label{fig:radsurfdens}} \end{figure*}

We have used the tilted ring parameters as described in
\cite{weldrake03} to derive a radial profile of the smoothed stellar
surface density. This is presented in Fig.~\ref{fig:radsurfdens}. The
profile does not start to turn over until the very outer parts of our
field at $|R| \ga 1700''$. The stellar densities reached there are
$1.43 \pm 0.38$ per 12$''$ pixel at the NW side, and $0.76 \pm 0.44$
per 12$''$ pixel at the SE side. This may indicate either a slighly
variable stellar Galactic foreground density, or alternatively, an
assymmetry in the extended stellar distribution of NGC 6822.  A
log-log plot of the same data shows no change in slope at larger
radii, indicating that the edge of the disk has not yet been reached.

Nevertheless, despite the possibly present small ``contamination'' by
NGC 6822 we will use the average surface density value of 1.1 stars to
define an ``edge'' to the stellar disk. This level is indicated in
Fig.~\ref{fig:surfdens} by the thick countour.  NGC 6822 covers almost
the entire field, only the most extreme eastern part can be said to
represent the field.  The extent of NGC 6822 is much larger to the NW
than to the SE.  This might tentatively be associated with the
(disrupting?)  presence of the NW cloud.

The possibility that the slightly higher stellar density in the
extreme NW is simply due to a gradient in the local Galactic stellar
foreground density ($\sim 1.2$ degrees lower Galactic latitude than
the extreme SE) is unlikely. The stellar foreground density as
measured by 2MASS varies only by $\sim 7$ per cent over this range in
Galactic latitude, much less than the factor of $\sim 2$ measured. The
asymmetric extent can also not be caused by selective foreground
reddening and extinction in the SE field. We will show in
Sect.~\ref{sec:reddening} that the reddening is higher in the NW,
accentuating the asymmetry even further.

With the present limited field of view it is difficult to draw more
firm conclusions, but it is plausible that the higher stellar density
observed in the NW is associated with the NGC 6822-NW Cloud system.
It would be very interesting to further investigate the wider field
around the cloud, looking for e.g.\ the presence of tidal streams.

\subsection{Foreground Population}

The CMD of the merged catalogue as shown in Fig.~\ref{fig:shallowdeep}
shows three distinct components. Firstly, we see the vertical ``blue
plume'' around $B-R \sim 0.25$ which consists mostly of young stars in
NGC 6822.  The concentration of stars around $B-R \sim 1.5$ and $m_B
\sim 25$, dubbed the ``red-tangle'' by \citet{gallart1}, contains
mainly old and intermediate age NGC 6822 stars. The third component, a
vertical band around $B-R \sim 1.3$ consists of Galactic foreground
stars. We will discuss the first two components in some detail later;
here we concentrate on the (removal of) foreground stars.

In order to study the stellar populations in NGC 6822 one ideally
wants to isolate the contribution of the foreground stars from the
CMD.  This is usually done in a statistical manner by comparing the
CMDs of the galaxy field with one of a nearby ``empty'' field (i.e.\
without galaxy stars) and removing at each location in the galaxy CMD
a number of stars that is proportional to the number of stars at that
location in the field CMD. This process is described and illustrated
for NGC 6822 in \citet{gallart1}.

The extent and depth of our catalogues poses some interesting
problems.  Firstly, a statistical subtraction can only be considered
for the deep catalogue as we do not have the equivalent of a shallow
field area for the shallow catalogue: its coverage is much smaller
than the extent of NGC 6822.

Secondly, the statistical method assumes that sizes of both fields and
the number and density of fields stars in them are roughly equal. If
one field is much smaller, and therefore contains significantly less
stars, one runs into sampling problems and small number statistics.
This, unfortunately, is the case here: in our observations only a very
small area can be regarded as ``field'' (and even there some
contamination by NGC 6822 is likely to be present). The usable
``field'' part in Fig.~\ref{fig:surfdens} (i.e.\ with stellar surface
density less than 1.1) measures $\sim 0.012$ square degrees and
contains $\sim 1000$ stellar objects. This should be contrasted with
the rest of the field which spans $\sim 0.53$ square degrees and
contains some 250 times as many objects. Clearly a statistical
foreground subtraction will be dominated by sampling effects in the
small field area.  

Even though a proper statistical subtraction is not possible, we can
still get an idea of the contamination by field stars and avoid the
problems of small number statistics by looking at the binned surface
densities of stars in the field and galaxy CMD diagrams.  To do so we
have binned the field and galaxy CMDs in $(B-R) \times m_B$ bins of
$0.25 \times 1$ mag and counted the numbers of stars in each bin.

\begin{figure*}[t!]
\plotone{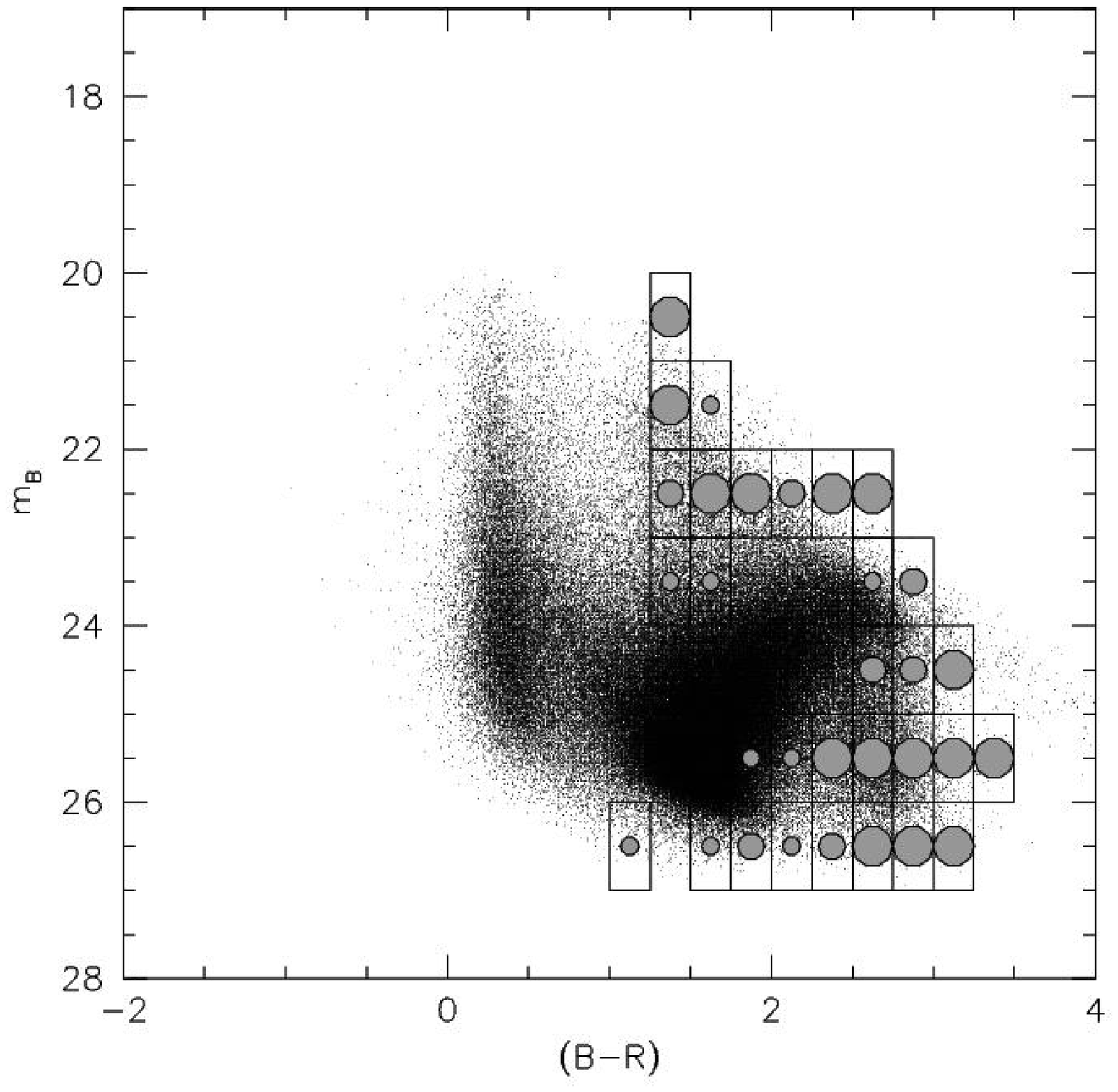}
\figcaption{CMD of the deep catalogue with degree of field
  contamination overplotted. Stars were counted in each CMD bin (as
  indicated in Figure), and compared after scaling (see text). The
  largest symbols indicate a field star density between 0.5 and 2.0
  times the galaxy star density, that is, the regions dominated by
  field stars.  Intermediate symbols indicate a field fraction between
  0.25 and 0.5.  The smallest symbols indicate a field fraction
  between 0.25 and 0.15. Areas without symbols have a field fraction
  less than 0.15, and can be considered to be uncontaminated by field
  stars.
\label{fig:fieldcount}} \end{figure*}

When normalized by the respective field sizes we expect for a pure
field population to find equal numbers of stars in equivalent bins in
both diagrams (modulo counting statistics).  Figure
\ref{fig:fieldcount} shows the ratios in each bin.  The largest
symbols mark the bins where the number of stars in the field CMD bin
equals the (normalized) number of stars in the equivalent galaxy CMD
bin to within a factor of two; these are thus areas
dominated by field stars. The smallest symbols mark the bins where the
number of stars in the field CMD bin is a quarter to a tenth of the
(normalized) equivalent number in the galaxy CMD bin.  Regions with
ratios less than 10 per cent (the rest of the CMD) can be safely said
to be dominated by NGC 6822 stars.

A comparison with Figs.\ 12 and 16 in \citet{gallart1} shows that our
method has been succesful in identifying CMD regions contaminated by
foreground stars. It has flagged the lower part of the plume of
Galactic stars around $(B-R) \sim 1.5$. The brighter part of this
plume is clearly visible in the shallow CMD.  Also clearly identified
as due to foreground stars is the region between $2 < B-R < 3$ and $27
< m_B < 24.5$.

\subsection{Distribution of populations}

In the following, when studying the stellar populations of NGC 6822 we
will only use those parts of the CMD that are deemed to contain only a
negligible component of foreground stars, i.e.\ the unmarked areas in
Fig.~\ref{fig:fieldcount}.

There are two distinct features visible in the uncontaminated CMD of
NGC 6822 (also described in \citealt{gallart3}).

$\bullet$ \emph{Blue Plume}. This is the area with $(B-R) < 0.75$
consisting mostly of main sequence (MS) and helium-burning Blue Loop
(BL) stars. In the absence of variable foreground extinction as found
towards NGC 6822, MS stars would occupy the blue side of the plume, BL
stars the red side. This area contains mostly young stars down to an
age of roughly 0.5 Gyr.

$\bullet$ \emph{Red-Tangle}. This area, with $1 \la (B-R) \la 1.75$
and $m_B \ga 23$, contains a mixture of old populations with ages
between $\sim 1$ and $\sim 10$ Gyr. It contains RGB stars, the fainter
AGB stars, and intermediate age BL stars.  The blue observing bands
used here are clearly not suited to disentangle the extended star
formation history (see \citealt{gallart1}).  However, studying the
red-tangle can give a good idea of the presence and importance of old
and intermediate age populations.

\begin{figure*}[t!]
\plotone{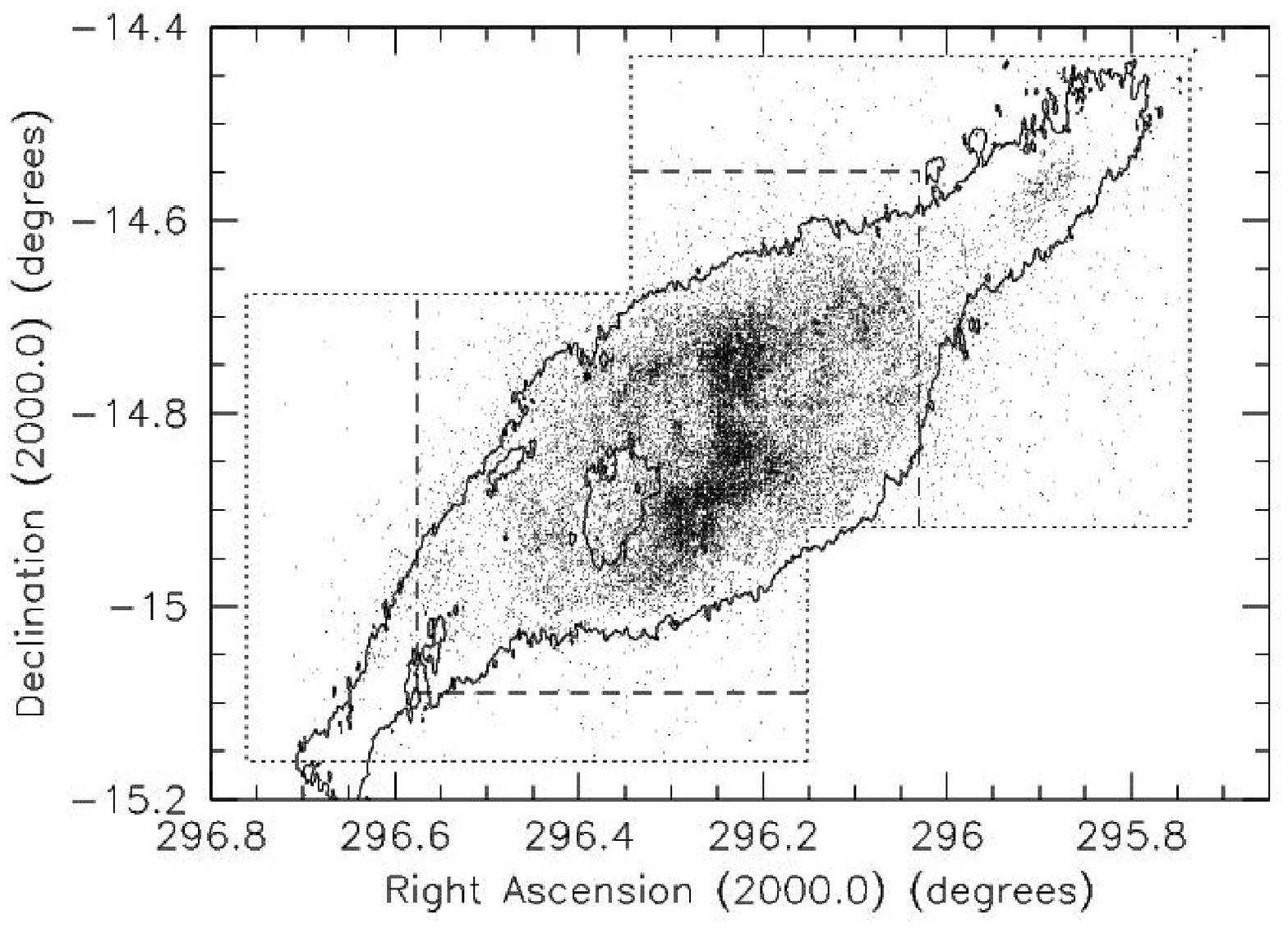}
\figcaption{Distribution of blue stars on the sky. Areas covered by
  the deep (short dashed box) and shallow (long dashed box) are
  indicated. Inside the area of the shallow catalogue we plot the blue
  stars from the merged catalogue. The contour indicates the edge of the \HI disk at $5
\cdot 10^{20}$ cm$^{-2}$ (not corrected for inclination).
\label{fig:plotblue}} \end{figure*}

The presence of a young stellar population outside the convential
optical extent of NGC 6822 is now well established and described in
\citet{deblokwalter03,komiyama03,letarte03}. The current analysis improves on 
all these studies by the deeper limiting magnitude.  Figure
\ref{fig:plotblue} shows the distribution of all stars $B-R <
0.75$. We show the blue stars from the combined catalog in the inner
part and those from the deep catalogue in the outer part.  We have
checked the outer field for saturated stars that should have been
incorporated in an equivalent shallow catalogue of the outer field,
but find there are very few of those present. The distribution in
Fig.~\ref{fig:plotblue} is thus representative of the true blue star
distribution, despite the lack of a shallow outer field catalog.

\begin{figure*}[t!]
\plotone{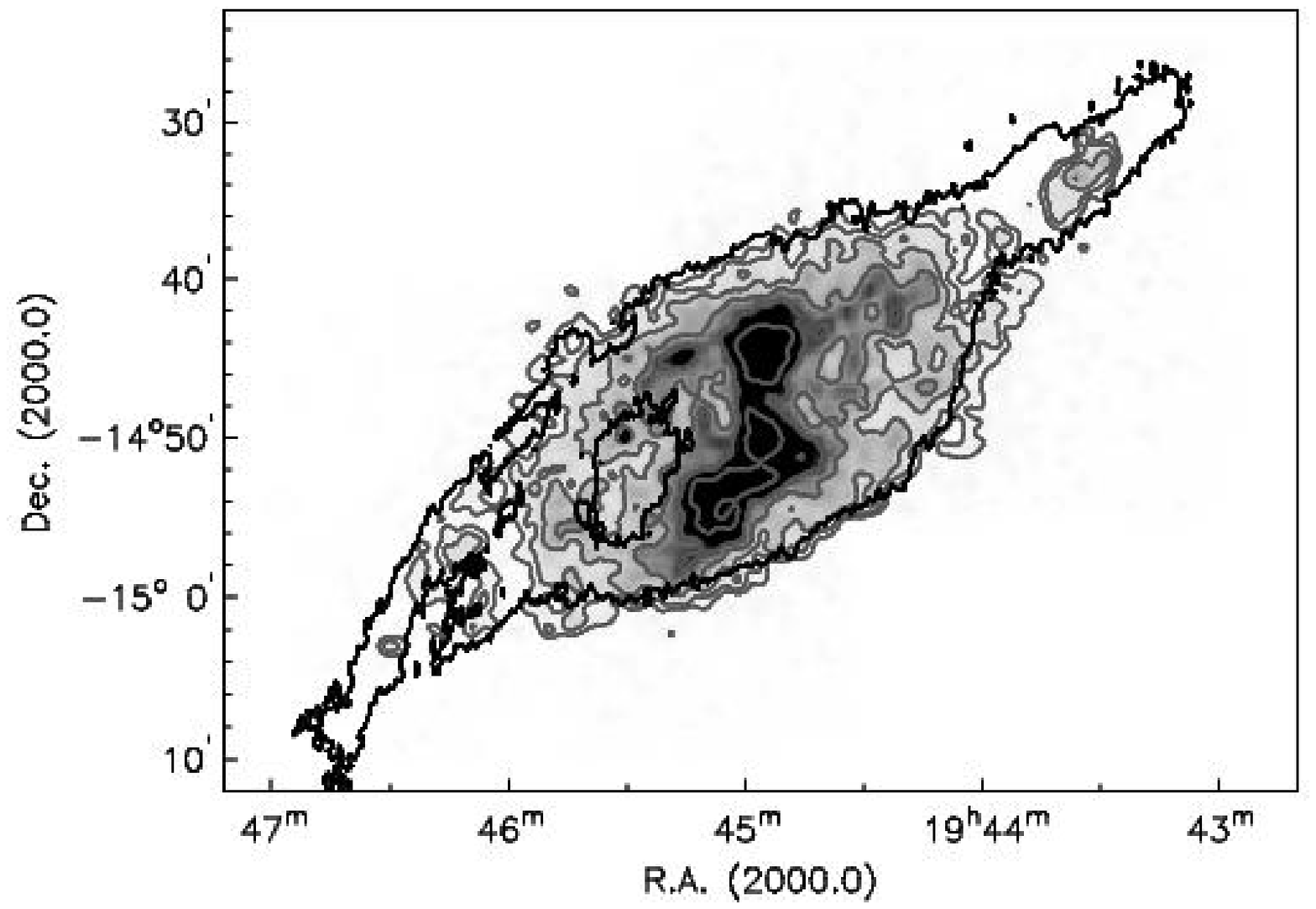}
\figcaption{Number density of blue stars measured in $12'' \times 12''$ 
boxes, smoothed to a final resolution of $48'' \times 48''$.
The grayscale levels run from 0 to 5 stars per smoothed $12''$ pixel.
The contour levels indicate 0.25, 0.5, 1, 2, 4 and 8 stars per pixel.
The contour indicates the edge of the \HI disk at $5
\cdot 10^{20}$ cm$^{-2}$ (not corrected for inclination).
\label{fig:bluedens}} \end{figure*}

At this point it is instructive to also show the surface density of
blue stars (Fig.~\ref{fig:bluedens}). This has been derived in a
similar way as Fig.~\ref{fig:surfdens}, this time only counting stars
with $B-R<0.75$ in $12'' \times 12''$ boxes. The resulting
distribution was smoothed to a resolution of $48''$.

Comparing Figs.~\ref{fig:plotblue} and \ref{fig:bluedens} a couple of
impressions from previous work are reinforced here.  Firstly, blue
stars are found out to the edge of the \HI disk. Secondly, the
distribution of blue stars is clumpy. The main stellar body of N6822
is easily recognizable, as well as the ``Blue Plume'' (the large
concentration of young stars to the SE of the central component)
described in \citet{deblokwalter03} (and not to be confused with the
CMD Blue Plume from \citealt{gallart3}).  Thirdly, the Super Giant
Shell (SGS) in the SE, described in \citet{deblokwalter00} is visible
as a slight underdensity in the stellar distribution. It is however
clear that because of the superior quality of the new data compared to
those of \citet{deblokwalter03} we can now actually detect a stellar
population in the hole (cf.\ \citealt{komiyama03}).

The nature of the NW cloud as a separate entity is again confirmed
with a clear underdensity of stars separating the cloud from the main
body of the galaxy.  The NW cloud will be discussed in more detail in
Sect.~\ref{NW}

\begin{figure*}[t!]
\plotone{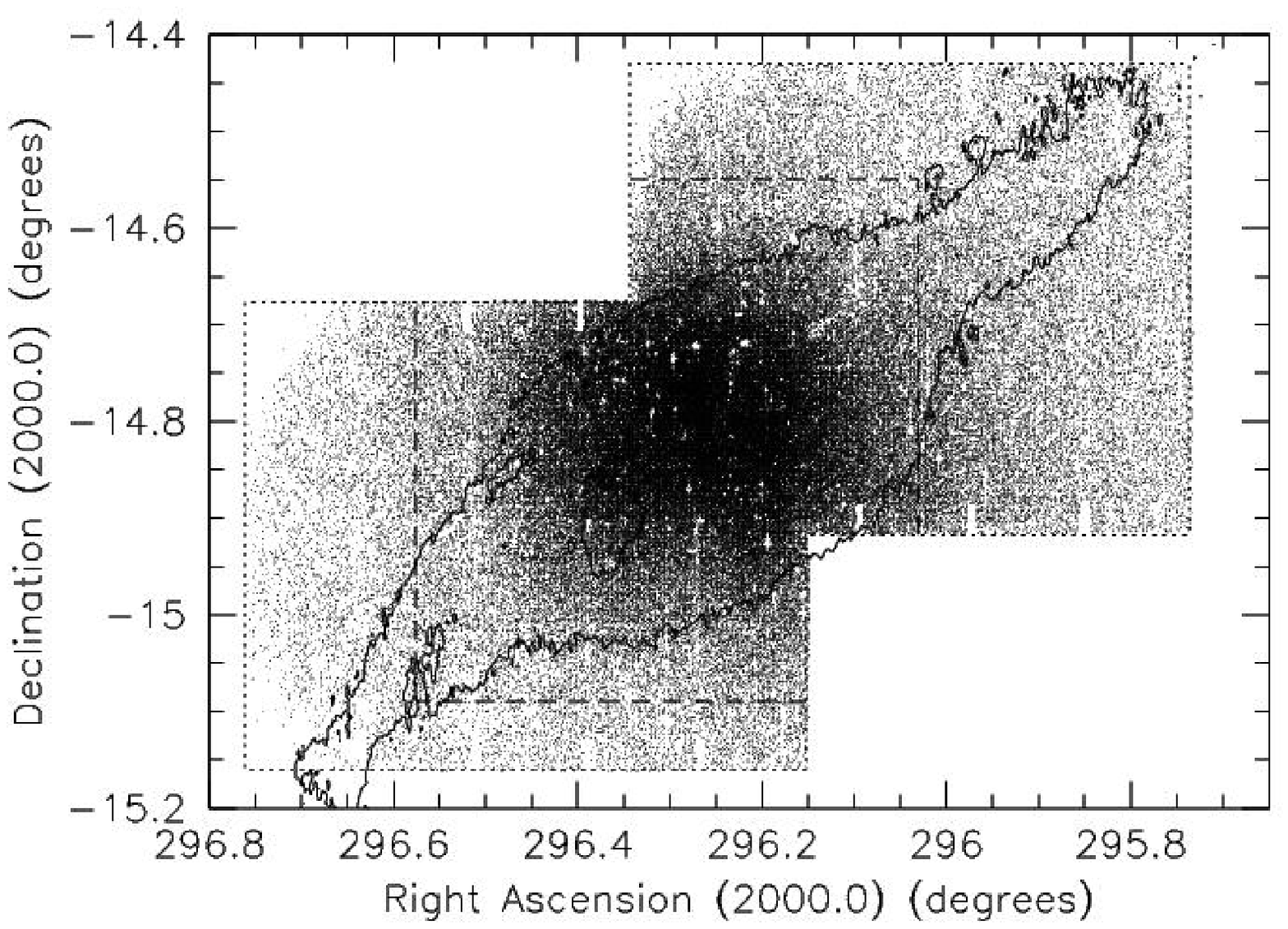}
\figcaption{Distribution of the old population on the sky.  Only stars
  in field uncontaminated areas of the CMD are plotted. The contour indicates the edge of the \HI disk at $5
\cdot 10^{20}$ cm$^{-2}$ (not corrected for inclination).
\label{fig:plotred}} \end{figure*}

The distribution of the old and intermediate population can be mapped
in a similar way. We consider the unaffected part of the CMD with
$(B-R)>1$.  Its distribution on the sky is shown in
Fig.~\ref{fig:plotred}.  Though the apparent distribution on small
scales is affected by the presence of bright foreground stars, as well
as bright stars within NGC 6822, it is clear that the distribution of
the old and intermediate population resembles that of the total
population shown in Fig.~\ref{fig:surfdens}.

\subsection{The colour of the disk and reddening corrections\label{sec:reddening}}

We can derive the average colour of the stellar disk in a similar way
as the number density.  We bin the distribution of stars on the sky in
$48''$ pixels, and compute the average colour of the stars in each
pixel, but only for those pixels containing more than 3 stars. For
cosmetic purposes the resulting images have been median-filtered using
a $3 \times 3$ pixels median filter.

\begin{figure*}[t!]
\plotone{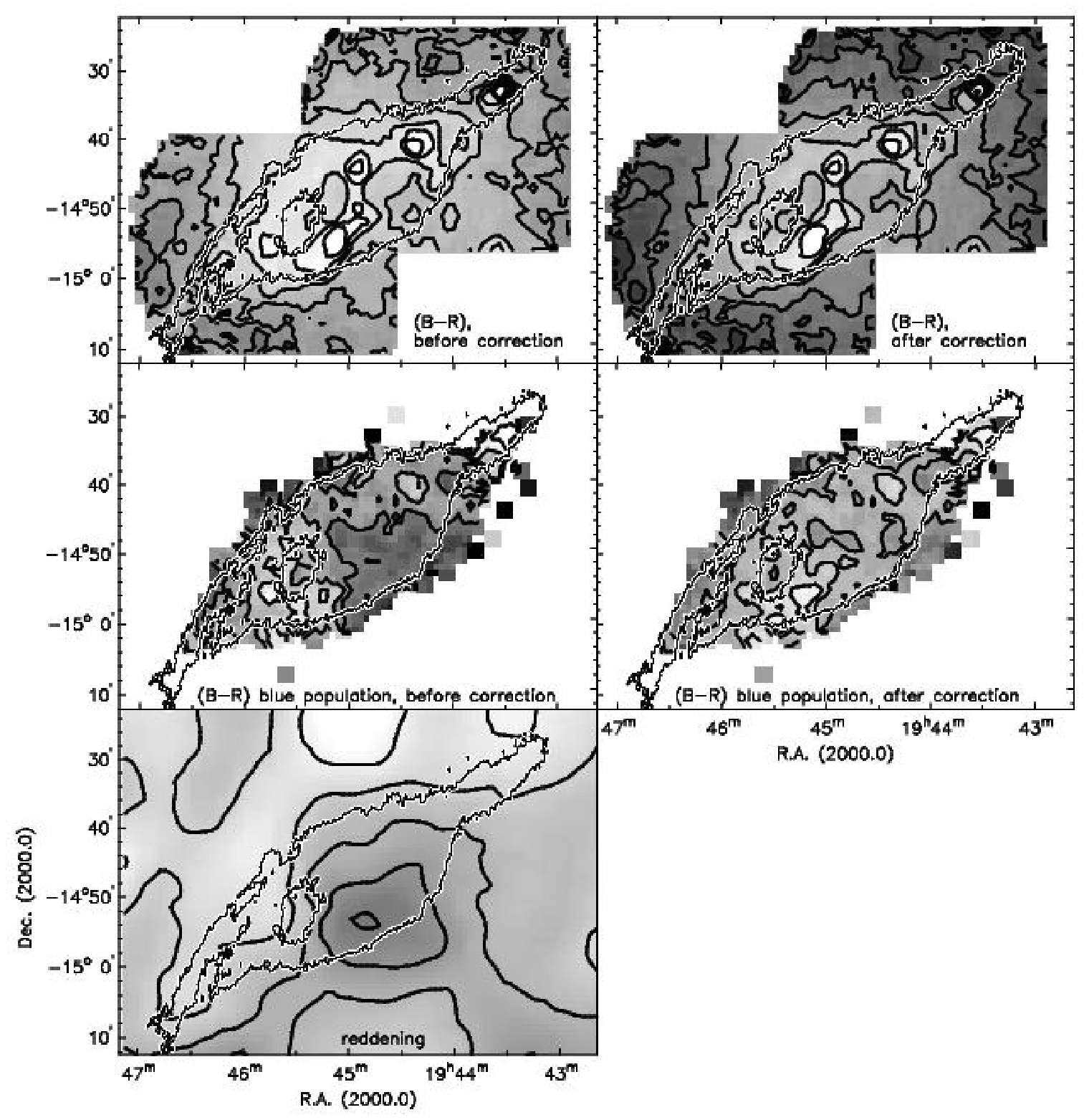}
\figcaption{Colour of the stellar disk. The top-left panel shows the
  colours of the field-uncontaminated disk population. The lowest
  (bluest) contour is $B-R = 1.3$. Contours increase in steps of 0.1
  from white to dark-gray. The top-right panel shows the same
  population but after correction for foreground reddening, as shown
  in the bottom-left panel.  The lowest contour is $B-R=0.85$, and
  contours increase in steps of 0.15 from white to dark-gray. The
  centre-left panel shows the colours of the field-uncontaminated blue
  star population only.  Contours run from $B-R = 0.3$ (white) to 0.55
  (dark-gray), in steps of 0.05.  The long contour running over the
  centre of the galaxy has value $B-R=0.45$.  The centre-right panel
  shows the same population but after correction for foreground
  reddening, as shown in the bottom-left panel. The lowest contour is
  $B-R = -0.05$, increasing in steps of 0.05 from white to
  dark-gray. The bottom-left panel shows the reddening values from the
  \citet{schlegel} maps. Contours run from $E(B-V) = 0.18$ (white) to
  0.26 (dark-gray) in steps of 0.02.   The thin white-black contour indicates the edge of the \HI disk at $5
\cdot 10^{20}$ cm$^{-2}$ (not corrected for inclination).\label{fig:colourdens}}
  \end{figure*}

The top left panel in Fig.~\ref{fig:colourdens} shows the average
apparent $B-R$ colour of the stellar disk.  The NW cloud jumps out as
being much bluer than the rest of the galaxy.  Also clearly visible as
blue regions are the areas associated with the Blue Plume and the main
optical bar. An additional blue region is observed at the SE edge of
the supergiant shell, corresponding with the locations of the newly
discovered H$\alpha$ regions. In fact, the distribution of the bluest
regions is strongly correlated with that of the H$\alpha$, as both trace
the most recent star formation. 

This unusual colour distribution is not intrinsic to the galaxy itself
though. Most of it results from the variable Galactic foreground
extinction towards NGC 6822.  \citet{massey95} already pointed out the
variable reddening based on measurements of spectra of hot stars in 5
fields near the star forming regions in the central northern part of
NGC 6822. They found reddening values $E(B-V)$ varying between 0.24
and 0.54.

The bottom panel of Fig.~\ref{fig:colourdens} shows the reddening
$E(B-V)$ towards NGC 6822 derived from the reddening maps by
\citet*{schlegel}. There is a strong correlation between the foreground
reddening and the observed colours of the disk, indicating that most
of the variation in the apparent colours of the disk is not intrinsic
to NGC 6822.

We can attempt to correct for the foreground reddening by converting
the \citet{schlegel} $E(B-V)$ reddening map to $E(B-R)$ units [where
$E(B-R) = 1.77\cdot E(B-V)$ for a $R_V = 3.1$ reddening law] and
subtracting this from the observed colour maps. The corrected colour
distribution is shown in the top-right panel of
Fig.~\ref{fig:colourdens}.

Most of the features in the colour map are still visible after
correction.  Unfortunately, at this stage we are unable to say whether
this is due to internal or external reddening. The resolution of the
\citet{schlegel} reddening map is less than that of the size of the
smallest displayed features, and it is very well possible that some
unresolved foreground reddening may give rise to the red edge,
especially as it occurs in the region where the largest reddening is
observed.

We can also look at the distribution of colours of the blue population
alone.  For this we have again created $48''$ resolution colour
images, but this time only including stars with $B-R<0.75$.  Limiting
the colour distribution to that of the blue population alone
emphasizes colour differences present among the young populations. The
colours are thus not ``dilluted'' by the presence of a prominent red
population.  These images, before and after reddening correction are
shown in the middle panels in Fig.~\ref{fig:colourdens}.  Before
correction, a large-scale reddening gradient is clearly visible as a
function of radius, however the distribution of disk colour is rather
unusual, with the western edge of the galaxy much redder than the rest
of the disk. After correction, the blue population in NGC 6822 shows
no strong colour gradient, consistent with the fact that (recent) star
formation is seen happening all over the disk.  The distribution of
the bluest features is intriguiging.  The NW cloud is still one of the
bluest parts of the galaxy. Another prominent blue feature is the
region around the Hubble {\sc x} star forming region
[$\alpha,\delta(2000.0) = 19^h45^m05.2^s, -14^\circ43'13'$], another
site of intense recent star formation.  Apart from these two regions,
all other bluest regions are found surrounding the supergiant shell in
the SE.  Intriguingly enough, we can now also see a red peak in the
centre of the supergiant shell, strongly suggesting the presence of a
colour gradient towards the edges of the hole.

In the following we will describe the NW cloud and the giant hole in
more detail.  Section~\ref{NW} will describe the properties of the NW
cloud, and argue that it is likely to be a separate dwarf galaxy that
has been captured by NGC 6822.  The supergiant shell will be discussed
in more detail in Sect~\ref{sec:SGS}.

\subsection{The stellar population of the NW cloud\label{NW}}

The discovery of the NW cloud as described in \citet{deblokwalter00}
was one of the unexpected results of the \HI observations described
there and in this paper. It was speculated by
\citet{deblokwalter00,deblokwalter03} that the cloud could be a
separate system that is currently in interaction with the main body of
NGC 6822, and responsible for the tidal arms in the SE. It could also
have triggered the star formation that would eventually lead to the
creation the large hole in the SE part of the main \HI disk. 

\begin{figure*}[t!]
\plotone{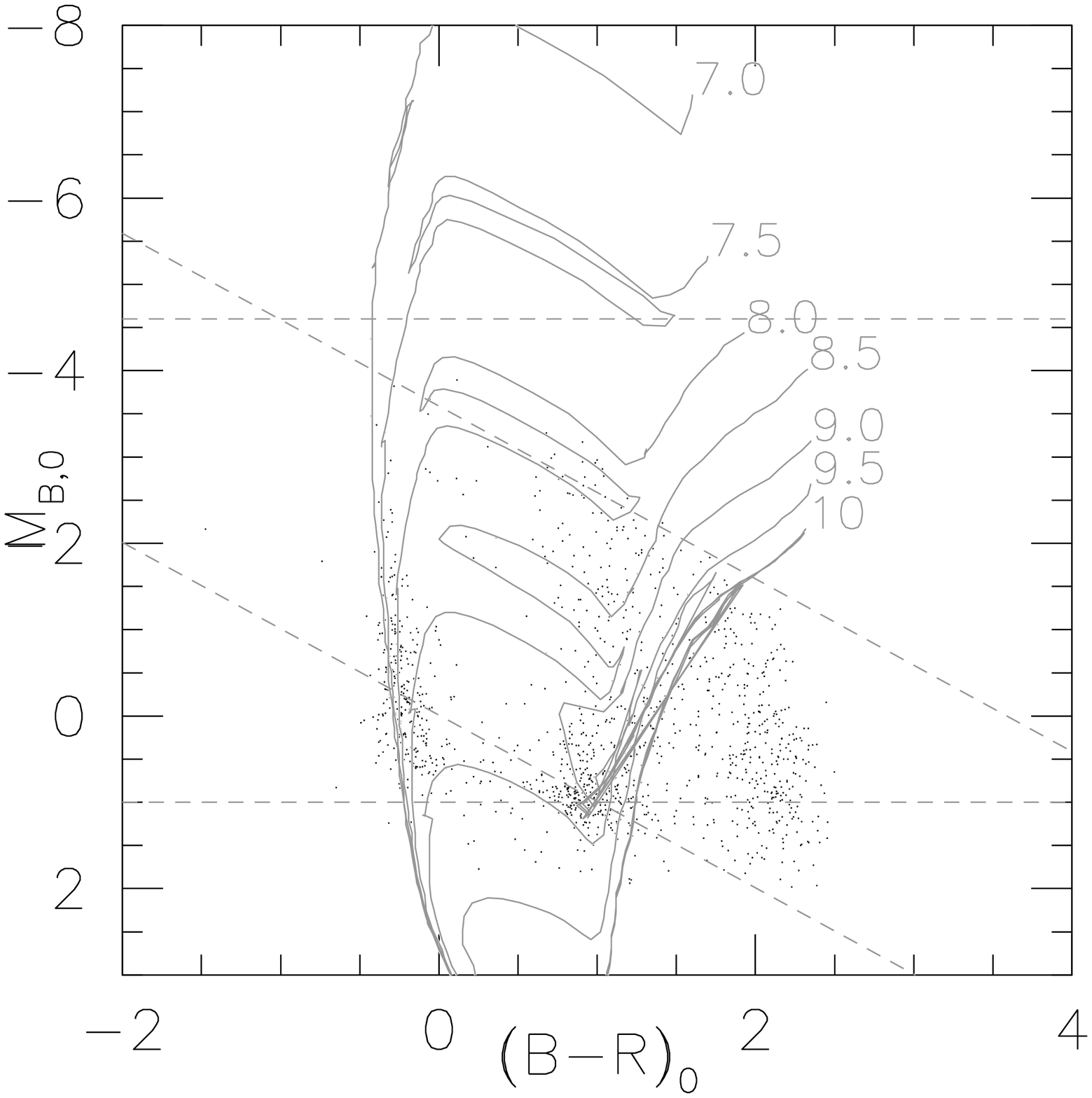}
\figcaption{CMD
  of the NW cloud. Several $Z=0.004$ isochrones from \citet{girardi00}
  are superimposed. The log(age) is indicated next to each
  isochrone. The dashed lines indicate saturation and completeness
  limits of the deep catalogue as indicated in
  Fig.~\ref{fig:shallowdeep}.
\label{fig:cloudCMD}} \end{figure*}

The cloud is amongst the bluest objects in the NGC 6822 system
(Sect.~\ref{sec:reddening}).  This indicates a significant amount of
recent star formation, which is confirmed by its CMD (see
Fig.~\ref{fig:cloudCMD}). Unfortunately, the cloud was not covered by
the shallow Subaru images, and we can therefore only measure stars up
to $M_B \sim -3$ in the colour range of the cloud main sequence
(cf.~Fig.~\ref{fig:shallowdeep}) before saturation sets in.

The uncontaminated part of the CMD of the NW cloud is dominated by a
well-defined main sequence, extending all the way up to the saturation
limit of the data. For a comparison with theoretical isochrones we use
the $Z=0.004$ model from \citet{girardi00}.  The brightest and bluest
stars are well described by a log(age/yr) $=7.0$ isochrone, though it
is likely that even younger stars are present, given the presence of
H$\alpha$ regions in the cloud (Fig.~\ref{fig:halfa}), and the
presence of a few saturated stars clustered in the same way as the
unsaturated stars.  It is thus very likely that the cloud has had
ongoing star formation over the past $10^7$ years, confirming previous
work.

Unfortunately the CMDs presented here do not allow us to make
quantitative statement on the SFH beyond 1 Gyr or
so. Fig.~\ref{fig:cloudCMD} shows there are some stars present
redwards of the MS that could be consistent with star formation at
log(age) $\sim 8.2$.

\begin{figure*}[t!]
\plotone{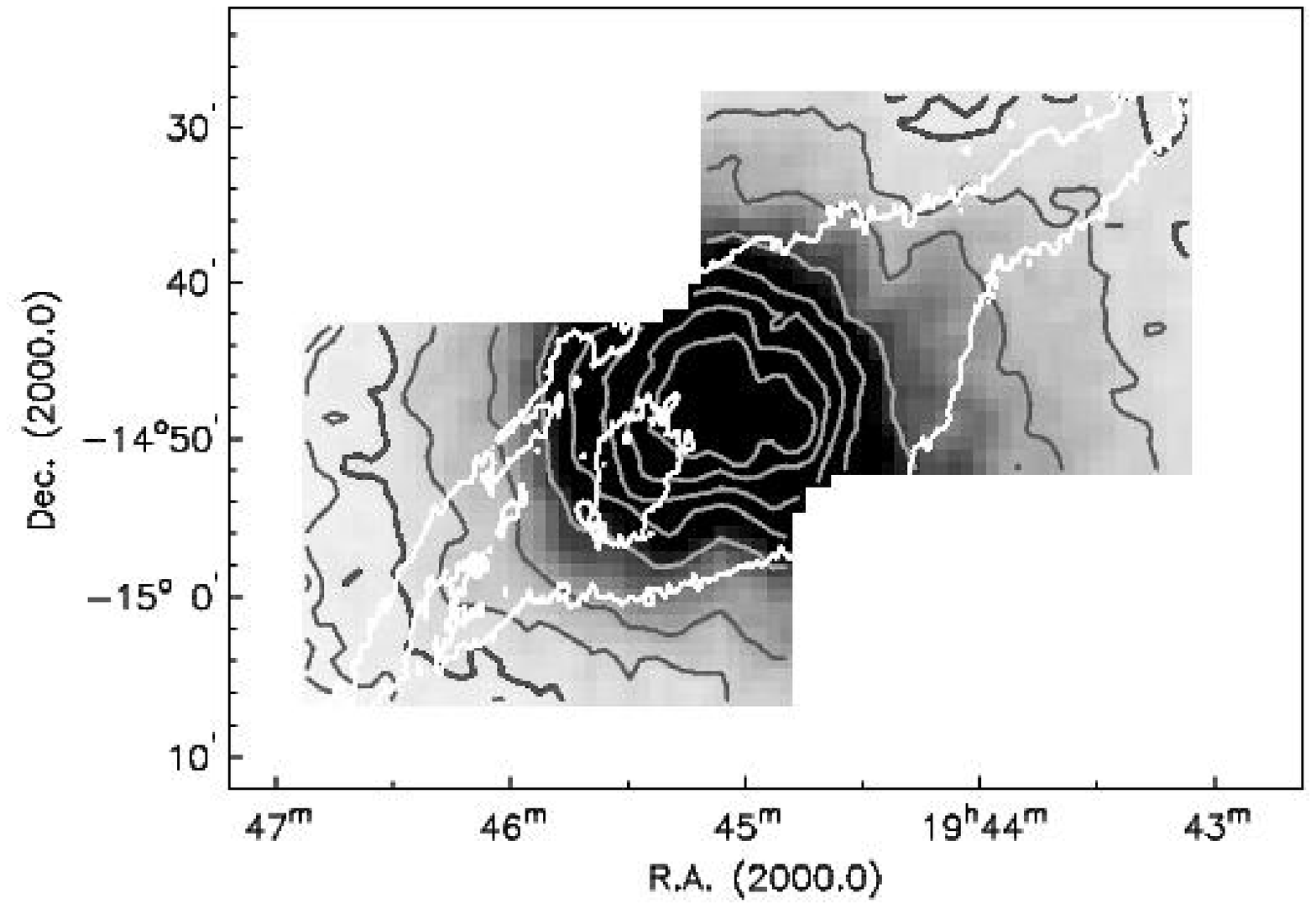}
\figcaption{Surface density of the old population measured in $12''
  \times 12''$ boxes, smoothed to a final resolution of $48'' \times
  48''$. Greyscale levels run from densities of 0 to 100. Contours
  with levels 2, 4, 8 (thick contour), 16, 32, followed by levels from
  80 to 280 in steps of 50. 
 The white contour indicates the edge of the \HI disk at $5
\cdot 10^{20}$ cm$^{-2}$ (not corrected for inclination).
\label{fig:reddens}} \end{figure*}

There are star present in the red-tangle, but a comparison between
Figs.~\ref{fig:bluedens} and \ref{fig:reddens} shows that whereas the
cloud is clearly present as an overdensity in the blue star
distribution, this is less obvious for the red population. The
clearest indication is a slight enhancement in the southern part of
the cloud.  This makes it likely that the cloud is ``old'', i.e., it
contains a stellar population older than 1 Gyr.

\subsubsection{The dynamics of the NW cloud}

If the cloud is indeed a separate dwarf galaxy we might expect it to
contain a significant amount of dark matter.  Fig.~\ref{fig:chans}
shows that the emission from the cloud is cleanly separated in
velocity from the main body of the galaxy, and we can thus examine the
velocity field of the cloud as shown in Fig.~\ref{fig:mom1}. The
kinematical major axis of the cloud has a different position angle
than that of NGC 6822. The main body of the galaxy has a kinematical
position angle between 103$^\circ$ and 140$^\circ$.  The cloud's
kinematical major axis varies between 70$^\circ$ and 90$^\circ$. 

The velocity gradient seen across the cloud can be interpreted as
either shear or rotation.  The velocity field of the cloud shows a
velocity range from $\sim -80$ to $\sim -105$ km s$^{-1}$, a range of
25 km s$^{-1}$.  A more detailed study of the data shows there is
cloud emission present from $-70$ to $-116$ km s$^{-1}$, but most of
the emission in the extended velocity range is of low surface
density. We will use 25 km s$^{-1}$ as a conservative estimate.  We
know from the the ages of the blue stars that the cloud has been
around for at least $5 \cdot 10^7$ years. A shear of 25 km s$^{-1}$
over that timescale implies a distance of 1.3 kpc, comparable to the
size of the cloud.  This means that, assuming that the shear remains
constant, the cloud will be completely disrupted over a similar
timescale, where it should be kept in mind that these timescales are
upper limits. We only observe the motion in the line of sight, and the
true disruption timescales would be much shorter.  Though it would be
possible to explain the kinematics of the cloud with pure shear, the
timescales involved are uncomfortably short.

A more likely explanation is that the velocity gradient is caused by
rotation.  Again, the signal of 25 km s$^{-1}$ is a lower limit, as we
do not know the space orientation of the cloud. The east-west diameter
of the cloud is $\sim 1.2$ kpc, and the associated (lower limit on
the) dynamical mass limit is therefore $2.2 \cdot 10^7$ $M_{\odot}$,
or $7.5 \cdot 10^{7}$ $M_{\odot}$ if we use the full velocity range
present in the data.

Adding up all detected stars in the NW cloud we derive a total
luminosity $M_B = -7.8$. This is again a lower limit, because any
unresolved as well as the brightest stars have not been included in
this total.  The addition of 2 or 3 stars with $M_B = -5$ (just above
the saturation limit) would increase the luminosity by 0.2 mag or so.
The number derived here is in good agreement with the value $M_B =
-8.5$ derived by \citet{komiyama03}.  If we assume $M/L_{*,B} = 1.0$,
we derive a stellar mass of $\sim 1.2 \cdot 10^5$ $M_{\odot}$.  If we
compare this number with the \HI mass $M_{HI}=1.4\cdot 10^7$ derived
in \citet{deblokwalter00} we see that the baryons in the NW cloud
primarily consist of gas: the cloud is extremely gas-rich. Taking the
numbers derived above at face-value we get $M_{HI}/L_B \sim 120$.  We
can attempt to correct for the contribution of the unresolved
population by comparing the total luminosity of all stars in our
catalog with the integrated magnitude $M_B = -15.85$ given in
\citet{hodge91}. We find that our catalogue underestimates the
luminosity by $\sim 1.6$ mag. Applying this same difference to the
cloud luminosity we get $M_B \sim -10.1$. This yields a ratio
$M_{HI}/L_B \sim 8$, which still makes the cloud an extremely gas-rich
object.   Comparing with the lower limit on the dynamical mass
derived above we see that the cloud must be dominated by dark matter
with $M_{\rm dyn}/M_{\rm vis} \sim 1.6$ as the absolute lower limit, but with
$M_{\rm dyn}/M_{\rm vis} > 4$ as a more likely value.
The estimates derived here are typical of dwarf irregular
galaxies. The most likely conclusion is thus that the NW cloud is a
separate galaxy.

\subsection{The Super-Giant Shell\label{sec:SGS}}

As noted above, Fig.~\ref{fig:colourdens} shows a distinct red peak in
the centre of the supergiant shell (SGS) in the SE part of NGC 6822,
strongly suggesting a colour gradient towards the edges of the hole.
The SGS was first described in \citet{hodge91} and later in more
detail in \citet{deblokwalter00}.  It measures around 1 kpc in
diameter, and is by far the largest structure observed in the \HI disk
of NGC 6822.  From dynamical arguments \citet{deblokwalter00} derived
an age of the SGS of $\sim 130$ Myr, assuming that the hole has just
stalled.  The required energy to create a hole of this size is
equivalent to that of $\sim 1000$ supernovae. \citet{deblokwalter03}
searched for a remnant population in the SGS but found none. Their
data were however only sensitive to stars with $M_B \sim -1$ or
brighter, and could thus make no statements on the presence of
populations older than $\sim 80$ Myr.  \citet{komiyama03} made a
initial examination of the distribution of blue stars in the SGS area
and concluded that there were fewer blue stars present in the SGS
compared to the rim of the hole.  Our analysis of these data reaches a
fainter limiting magnitude and thus allow to derive more definitive
conclusions regarding the evolution of the SGS. Similar structures
have been in the HI distribution of many nearby dwarf galaxies when
observed at sufficient resolution (e.g., Holmberg II, Puche et
al. 1994, IC\,2574, Walter \& Brinks 2000; Holmberg I: Ott et
al. 2001; DDO 47: Walter \& Brinks 2001, SMC: Kim et al.\ 2003, LMC:
Staveley--Smith et al.\ 1997).

To explain these giant holes in the \HI disks of galaxies, several
explanations have been put forward. The most obvious one is that the
effects of effects of stellar winds and supernovae can create a hole
or bubble in the ISM (e.g., Tenorio--Tagle \& Bodenheimer 1988).  The
compression at the rim of the resulting bubbles can then lead to
secondary star formation, thus enlarging the hole even
further. Alternative explanations are the impact of HVCs, and
gamma-ray bursts (e.g., Kamphuis et al.\ 1991, Loeb \& Perna 1998,
Efremov et al.\ 1999). While these processes all inject energy into
the ISM, the time-scales and resulting spatial distributions differ.
E.g., \citet{vorobyov} show that effects on the gas disk and the
corresponding observational signatures are all distinctly different,
and based on comparison with observations conclude that star formation
is the most likely cause. And indeed, there are some cases in which
remnant clusters in kpc--sized holes have been identified, e.g., in
IC\,2574 (Walter et al.\ 1998, Stewart \& Walter) --- however it
should be noted that most past searches for remnant clusters in HI
superstructures have not been successful (e.g., Holmberg II: Rhode et
al.\ 1999, LMC-4: Braun et al.\ 1997).

One of the main arguments against the SF hypothesis has always been
that holes are also found in the outer disks of galaxies where
(usually) no stars or signs of star formation are observed.  If NGC
6822 is at all typical, then this argument no longer holds: its stars
are found all the way to the edge of the HI disk. This would imply
that closer inspection of the holes in other galaxies at similar
column densities should show a stellar population in many of them
(cf.\ \citealt{ferguson98}).

If the SGS is indeed created by star formation, we would expect a
radial age gradient in the stellar population (or, at least, an age
difference), with the most recent star formation found at the rim of
the hole.  Here we first investigate the radial distributions of the
various ISM and stellar components as a function of distance from the
centre of the hole. As centre we take the centre of the hole in the
\HI distribution.  An ellipse fit to the outer rim yields orientation
parameters PA $= 158^{\circ}$ and an axis ratio $b/a = 0.75$. The
centre was found to be at $\alpha(2000.0) = 19^h54^m30^s,
\delta(2000.0) = -14^{\circ}53'40''$. We will adopt these values for
the rest of our analysis.

\begin{figure*}[t!]
\plotone{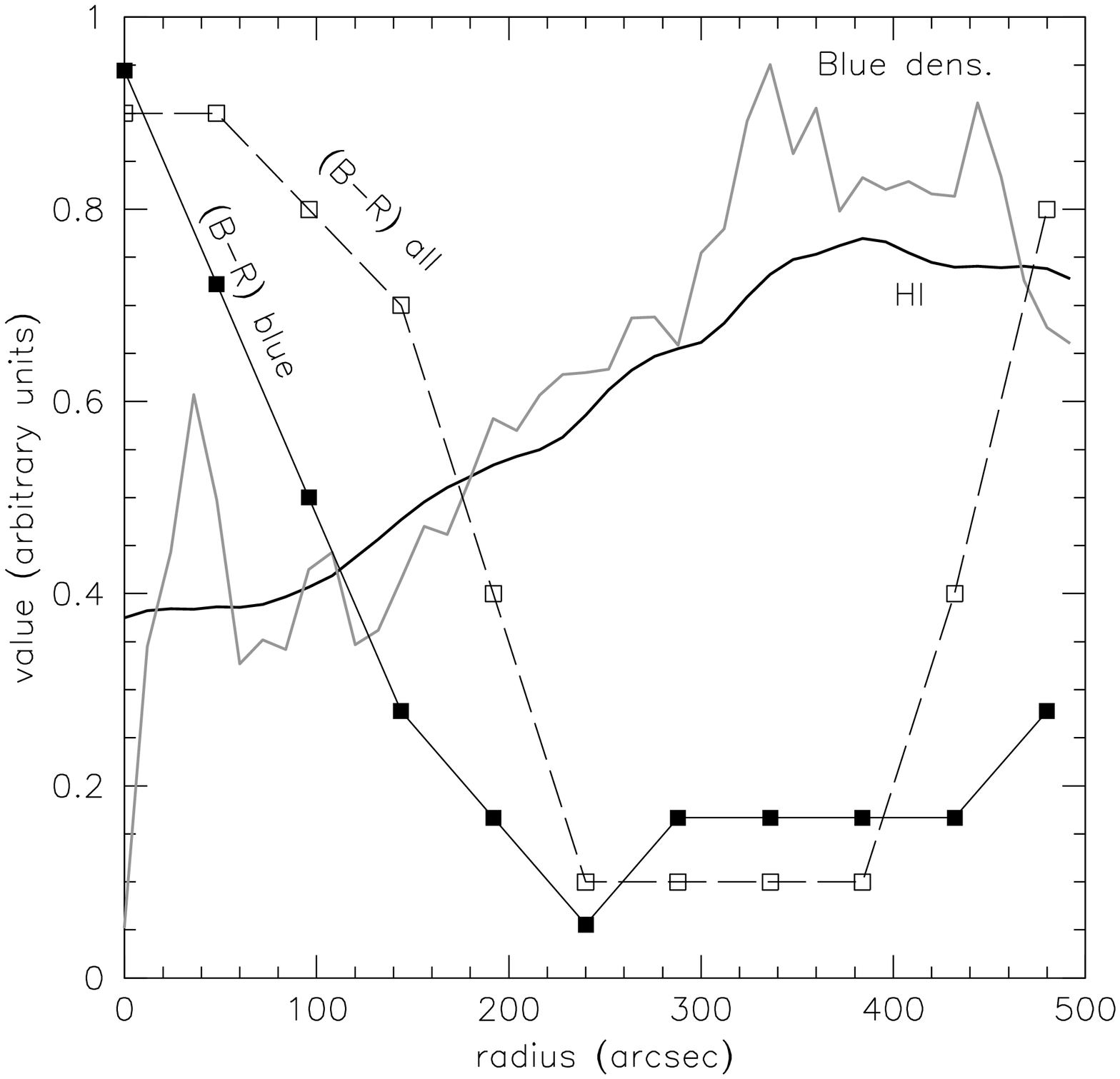}
\figcaption{Radial profiles with respect to the centre of the SGS.
  Shown are the \HI distribution (thick full line) and the density of
  blue stars (thick grey line). Also shown are the colour of the blue
  stars (full line with filled squares) and the colour of the total
  population (dashed line with open squares). The units along the
  vertical axis are arbitrary.
\label{fig:holerad}} \end{figure*}

Figure \ref{fig:holerad} shows the radial distributions of the H{\sc
i}, as well as those of the number densities and colours of the
stellar population.  It is immediately obvious that the number density
of blue stars closely follows that of the H{\sc i}. We also see a very
pronounced colour gradient from red colours in the centre of the hole,
getting bluer towards the rim.

To show that these radial trends are not an artefact of the azimuthal
averaging procedure, we show the distribution on the sky of these
components in Fig.~\ref{fig:hole2d}.  Superimposed on the various
panels are two ellipses with radii of $250''$ and $430''$,
corresponding with, respectively, the radius where the blue $B-R$
gradient flattens off, and the peak in the radial H$\alpha$
distribution.

In the case of the blue stars, the radial trend is clearly dominated
by the presence of the main body of NGC 6822 to the NW of the
SGS. However, note that this does not explain the entire gradient. At
azimuthal angles away from the main body of NGC 6822, we find that the
maxima in the HI distribution only occur between the two ellipses with
radii of 250$''$ and $430''$, as described above. The same applies to
the H$\alpha$: though few in number, the H$\alpha$ regions directly
surrounding the hole are only found between the two ellipses. This
also applies to the blue star distribution: the highest densities
occur between the two ellipses. The gradients measured are thus not
artifically introduced by the nearby presence of the main body of NGC
6822.

We can thus divide the SGS area in three zones: 1) the central zone at
$R<250''$ where there is no current star formation, and where there is
a color gradient with the reddest stars found in the centre; 2) the
annulus at $250''<R<430''$ where current and very recent star
formation is found.  This is the zone where the effects of the hole
are currently impacting on the ambient ISM; 3) the zone with $R>430''$
that is unaffected by the SGS.

We can now use CMDs to see whether the reddening gradient described
above translates into actual age gradients. A comparison with the
reddening map in Fig.~\ref{fig:colourdens} shows that it is extremely
unlikely that the gradient is caused by differential foreground
reddening.

We have divided the area within the $R=430''$ ring in three annuli
roughly equally spaced in radius: a central region $R<170''$ chosen to
include the reddest parts of the SGS; a transition annulus
$170''<R<300''$; and an outer annulus with $300<R<430''$ covering the
bluest area with very recent star formation. The borders between these
annuli are indicated in the top-left panel of Fig.~\ref{fig:hole2d}.

\begin{figure*}[t!]
\plotone{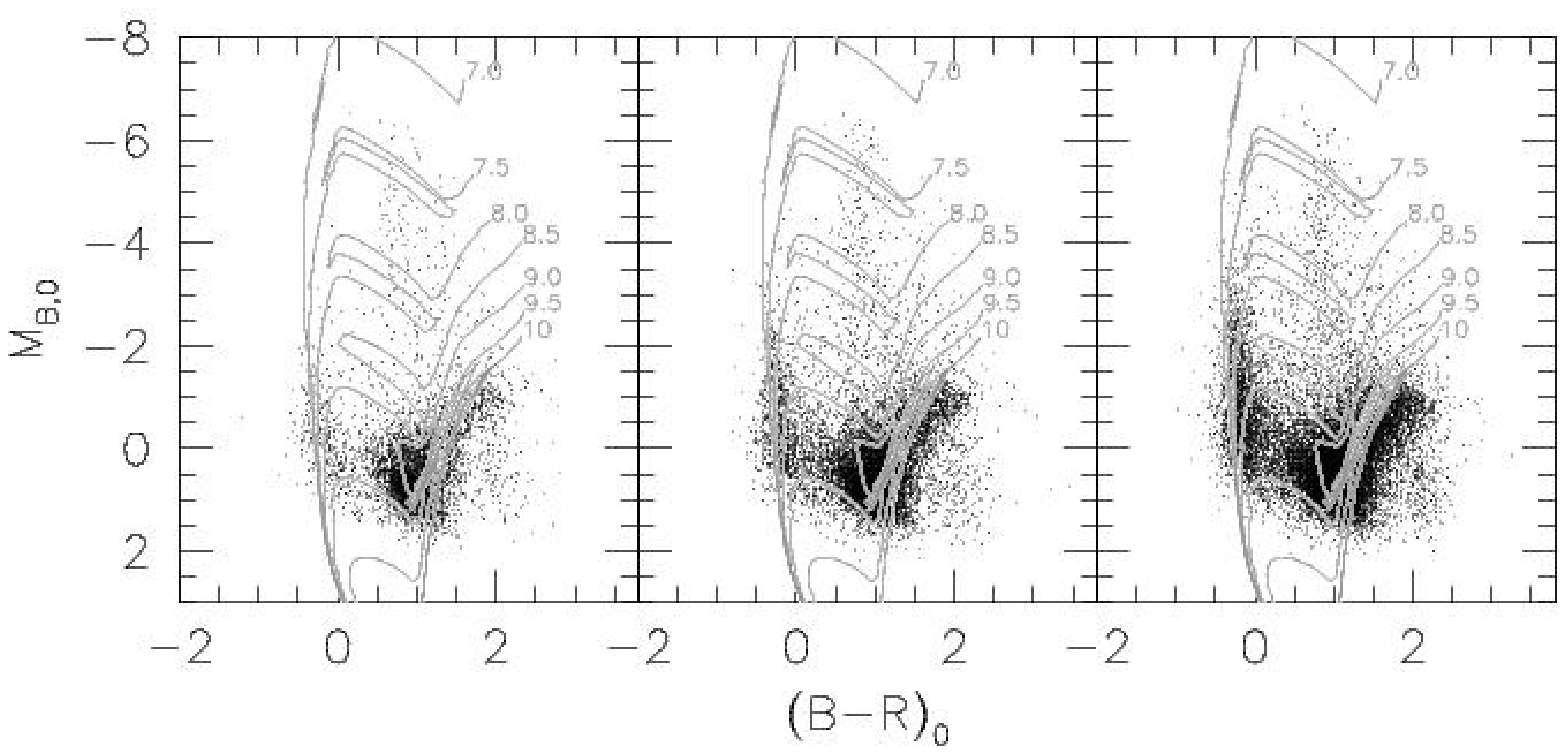}
\figcaption{CM
  diagrams of annuli in the SGS. Left panel shows the CMD of the star
  within the $170''$ semi-major axis ellipse shown in the previous
  figure. The central panels shows stars between semi-major axes
  $170''$ and $300''$. The right panel shows the stars in the annulus
  with semi-major axes between $300''$ and
  $430''$.  \label{fig:holeCMD}} \end{figure*}

The CMDs of these three regions are displayed in
Fig.~\ref{fig:holeCMD} with the usual isochrones overlaid.  We have
used the combined deep and shallow catalog, meaning that at the $B-R$
colour of the NGC 6822 Main Sequence we can detect stars as bright as
$M_B \sim -7$ without being affected by saturation effects
(cf.~Fig.~\ref{fig:shallowdeep}).

The left-hand diagram in Fig.~\ref{fig:holeCMD} shows the CMD
corresponding to the central area of the hole. The brightest and
bluest stars are located well away from the saturation limits of the
data so that our age conclusions are not affected by these limits. The
positions of the bright, blue stars in the CMD are best fitted
the log(age) $=8.0$ isochrone. A slightly younger age, such as
log(age)$= 7.7$ or $7.8$ is also possible but not preferred. The
log(age) $=7.5$ contour is definitely not a good description of the
data.

In the centre CMD, corresponding with the transition annulus in
Fig.~\ref{fig:hole2d}, the bluest and brightest stars are best fitted
with the log(age) $=7.5$ isochrone.  Here we are still well away from
the saturation limits of the data so that again the conclusions are
not affected in that way.

The right-hand panel shows the CMD of the outermost annulus,
which contains most signs of recent star-formation. This shows itself
in the CMD, where the brightest NGC 6822 stars are best fitted
with the log(age) $= 7.2$ isochrone. There are stars present along the
log(age) $=7.0$ isochrone as well, and stars along the brightest part
of this isochrone (as well as those along younger isochrones) will
have been affected by saturation effects, so in this part of the rim
of the galaxy may prefer a younger age as well. Certainly the presence
of H$\alpha$ is evidence for star formation on timescales less than
$\sim 10^7$ year.

In this sequence of CMDs we thus see age differences in the hole
region: The stellar population of the inner part of the hole can be
described with an age between $5\cdot 10^7$ and $10^8$ yr, while on
the rim of the hole recent star formation is still going on.

This conclusions puts the age of the hole, or rather the age of its
young stellar population at between $5\cdot 10^7$ and $10^8$ yr. This
is slightly younger than the dynamical estimate of $1.3 \cdot 10^8$ yr
(or log(age)$ \simeq 8.1$), however, the left panel in
Fig.~\ref{fig:holeCMD} shows that the stellar and dynamical age
estimates are consistent with each other.

The properties of the SGS in NGC 6822 are thus consistent with it
being formed by the processes of star formation and stellar
evolution. It also suggests possible explanations as to why searches
for a progenitor population in shells in other galaxies have not
always been succesful.  Firstly, in many cases the progenitor
population was expected to be in the form of star \emph{clusters}. In
the SE part of NGC 6822 where the SGS resides, star clusters are
conspicuous by their absence. The stars are distributed almost
homogeneously throughout this part of the galaxy. With a dispersion of
$\sim 1$ km s$^{-1}$ a star cluster will after $10^8$ yr have expanded
to a size of $\sim 0.1$ kpc, and may be difficult to distinghuish from
the background population.  The background itself is also difficult to
detect: the total luminosity of stars within the $R=170''$ radius is
$m_b = 14.1$. After correction for Galactic extinction this translates
into a surface brightness $\mu_B = 25.5$ as observed, or $\mu_B =
26.2$ when corrected to face-on. It is clear that if we had not been
able to resolve the stellar population of NGC 6822 we would have have
great difficulty detecting the unresolved progenitor population of the
SGS.

Though it is dangerous to extrapolate from a sample of one, the
temptation is to conclude that no exotic explanations are necessary to
explain the presence of SG shells in galaxies.  The reason that there
have not been many clear identifications of progenitor populations is
that by the time the shells have reached the large size observed, the
progenitor clusters have diffused enough to make them
indistinghuishable from the background. If this background is
unresolved then its low surface brightness makes a detection difficult
for any but the nearest galaxies.

\section{Summary}

We have presented a comprehensive study of the stellar population and
the interstellar medium in NGC\,6822, one of the most nearby dwarf
galaxies.  Its nearby location and our \HI (ATCA/Parkes) and optical
(Subaru/INT) observations allow for the first
time to perform a detailed comparison of respective distributions of
the \HI and the stellar populations. The main results are briefly
summarised in the following:

$\bullet$ Multi--array/mosaicked \HI observations obtained at the ATCA
(zero-spacing corrected using Parkes data) show a stunning morphology
in the H{\sc i}, including the presence of \HI holes of various sizes and a
detached cloud in the north--west. Azimuthally averaging the HI
distribution gives an 'edge' to the \HI disk at
5$\times$10$^{20}$\,cm$^{-2}$ --- the total \HI mass of NGC\,6822 is
1.34$\times$10$^{8}$\,M$_\odot$.

$\bullet$ Deep H$\alpha$ imaging (reaching one magnitude deeper than
previous observations) reveals the presence of a stunning filamentary
network which covers almost the entire central disk of NGC\,6822. Our
maps also reveal the presence of previously unknown HII regions in the
outskirts of NGC\,6822 and some coincident with the likely companion
galaxy. The total H$\alpha$ luminosity of NGC\,6822 is $(2.0 \pm 0.4)
\cdot 10^{39}$ erg s$^{-1}$.

$\bullet$ The combination of shallow and deep Subaru multi-band
imaging allows us, for the first time, to derive an (almost) complete
census of the stellar population in NGC\,6822 above magnitudes of
$(m_B,m_R,m_I) = (25.6, 24.1, 23.4)$. Our cross-correlated catalog of
all stars comprises some $4\times10^5$ objects.  Even though the
stellar population can be traced out to radii of 0.6$^\circ$ we
conclude from our star counts that the optical edge of NGC\,6822 has
not yet been reached even in the wide-field Subaru observations.

$\bullet$ The old and intermediate population of stars is much more
extended than even the \HI disk. This distribution is not symmetric but
more extended towards the north-west, i.e. where the companion galaxy
is situated. In sharp contrast, the distribution of the young, blue
stars, closely follows the distribution of the \HI disk all the way out
to the \HI `edge' and displays a highly structured morphology.

$\bullet$ The companion galaxy in the north--east of NGC\,6822 shows
evidence for an underlying older population (through the colors and
its CMD) --- the current star formation activity, although likely
being triggered by the interaction with NGC\,6822 is not the first SF
episode in this object. From the \HI dynamics, \HI mass
($M_{HI}=1.4\cdot 10^7$) and the optical luminosity ( $M_B = -7.8$) we
derive $M_{\rm dyn}/M_{\rm vis} > 4$ for the companion.

$\bullet$ We show that the properties of the large HI hole (the most
prominent structure seen in the HI distribution) are consistent with
it having been created by past star formation activity. Hhis is
supported by the color gradient towards the edge of the hole (red in
the centre, blue on the rim) and CMDs obtained at various radii. We
speculate that the cluster(s) responsible for the creation have been
dispersed -- this may also explain some of the unsuccessful searches
for remnant clustes in superstructures of other dwarf galaxies.

In summary, NGC\,6822 came up with many suprises -- the structured
distribution of the young blue stars throughout the HI disk, the very
extended diffuse stellar halo reaching far outside the HI disk, the
presence of star formation all across the HI disk, the likely creation
of the huge HI hole by star formation. These results could only be
obtained because of the small distance of NGC 6822, enabling the
stellar population to be resolved and the ISM to be observed at
sub-kpc scales.  All of the above results would have been impossible
to obtain if NGC 6822 had been more distant (e.g., eight times further
away at the distance to the M\,81 group). This raises the question on
whether the results found here may also apply to other dwarf galaxies
and in this sense, NGC\,6822 gives us some interesting new
perspectives on the properties of dwarf galaxies in general. However,
clearly, future similar in-depth, high-quality observations of other
nearby dwarf galaxies are mandatory, though challenging, in order to
show that NGC\,6822 is not exceptional, but rather a typical dwarf
galaxy.

\begin{acknowledgements}
We like to thank Phil Massey and the Local Group Survey group for
making their reduction and analysis scripts available on-line. These
were of immense value to us in the initial stages of reducing the
Subaru data.  This research has made use of the USNOFS Image and
Catalogue Archive operated by the United States Naval Observatory,
Flagstaff Station (http://www.nofs.navy.mil/data/fchpix/).
\end{acknowledgements}

\clearpage

\begin{deluxetable}{lrrrrr}
\tablewidth{0pt}
\tablecaption{Beam sizes and noise levels\label{tab:noises}}
\tablehead{\colhead{}&\colhead{SD}&\colhead{B96}&\colhead{B48}&\colhead{B24}&\colhead{B12}}
\startdata
beam major axis $('')$ &1002& 349.4 & 174.7  & 86.4  & 42.4 \\
beam minor axis $('')$ &1002& 96.0 & 48.0 & 24.0 & 12.0 \\
pixel size $('')$&240&32.0 &16.0 & 8.0 & 4.0 \\
$\sigma_{\rm channel}$ (mJy beam$^{-1}$)& 80&13.2 & 8.5 & 5.3  & 3.9  \\
$\sigma_{\rm HI}$ ($10^{19}$ cm$^{-2}$) & 0.01& $0.07$ & $0.18$ & $0.45$ & $1.3$  \\
$S$ (Jy km s$^{-1}$) & 2266 & 2020 & 2050 & 1917 & 1909 \\
$M_{HI}$ $(\times 10^8$ \Msun) & 1.34 & 1.19 & 1.21 & 1.13 & 1.13 \\
\enddata
\tablecomments{For the interferometric data the noise levels refer to fully mosaicked central parts of the cubes. A channel separation
of 1.6 \kms\ is used for all data.}
\end{deluxetable}

\begin{deluxetable}{cll}
\tablewidth{0pt}
\tablecaption{Pointing positions ATCA observations\label{tab:atcapointings}}
\tablehead{\colhead{Pointing} & \colhead{$\alpha(2000.0)$} & \colhead{$\delta(2000.0)$}\\&\colhead{$\phn^{h}\phn^m\phn^s$}&\colhead{$\phn^{\circ}\phn^m\phn^s$}}
\startdata
1 & 19 43 00 & --14 21 30 \\
2 & 19 43 00 & --14 41 06 \\
3 & 19 44 09 & --14 50 54 \\
4 & 19 45 18 & --15 00 52 \\
5 & 19 46 27 & --15 10 30 \\
6 & 19 46 27 & --14 50 54 \\
7 & 19 45 18 & --14 41 06 \\
8 & 19 44 09 & --14 31 18 \\
\enddata
\end{deluxetable}
 
\begin{deluxetable}{lrccl}
\tablewidth{0pt}
\tablecaption{ATCA observing dates\label{tab:atcaconfig}}
\tablehead{\colhead{Array} & \colhead{Date} & \colhead{Pointing}&\colhead{Time}&\colhead{Baselines}\\ & & & \colhead{on source [min]}&\colhead{[metres]}}
\startdata
375  & 27 Jun 1999 & 1,2,3,4,5,6,7,8& 544& 30.6, 61.2, 91.8, 122.5, \\
     &              &        &     & 183.7, 214.3, 244.9, 275.5,\\
     &              &        &     & 336.7, 459.2, 5510.2, 5755.1,\\
     &              &        &     & 5785.7, 5846.9, 5969.4\\[3pt]
750D & 27 Jul 2000  & 1,2,3,8&  576& 30.6, 107.1, 183.7, 290.8,\\
750D & 25 Jul 2000  & 4,5,6,7&  601& 398.0, 428.6, 581.7, 612.3\\
     &              &        &     & 688.8, 719.4, 3750.0, 3857.1,\\
     &              &        &     & 4040.8, 4438.8, 4469.4\\[3pt]
1.5A & 1 Jan 2000 & 3,8& 574& 153.1, 321.4, 428.6, 566.3,\\
1.5A & 2 Jan 2000 & 4,7& 591& 719.4, 750.0, 887.7, 1040.8,\\
1.5A & 3 Jan 2000 & 5,6& 447& 1316.3, 1469.4, 3000.0, 3428.6,\\
1.5A & 4 Jan 2000 & 1,2& 568& 3750.0, 4316.3, 4469.4\\[3pt]
6A & 9 Mar 2000 & 3& 551&  336.7, 627.6, 872.4, 1086.8,\\ 
6A & 10 Mar 2000 & 1& 536& 1423.5, 1500.0, 1959.2, 2295.9, \\
6A & 11 Mar 2000 & 4& 563& 2586.8, 2923.5, 3015.3, 3352.0,\\
6A & 12 Mar 2000 & 2& 572& 4438.8, 5311.2, 5938.8\\[3pt]
6D & 15 Mar 2000 & 8& 555& 76.6, 367.4, 795.9, 1163.3,\\
6D & 16 Mar 2000 & 7& 599& 1285.7, 1362.3, 2081.6, 2158.2,\\
6D & 17 Mar 2000 & 5& 567& 2449.0, 2525.6, 3352.0, 3428.6,\\
6D & 18 Mar 2000 & 6& 585& 4714.3, 5510.2, 5877.6\\
\enddata
\end{deluxetable}

\begin{figure}[t]
\epsscale{0.8}
\figcaption{Channel maps of the B12 data cube. Velocities are
  indicated in the top-left corners of each sub-panel. The grayscales
  run from 0 mJy beam$^{-1}$ (white) to 90 mJy beam$^{-1}$ (black). Contours in the
  top-left channel indicate (from the inside out) where the
  sensitivity has dropped to 90, 80 and 60 per cent of the central
  value. The channel spacing is 1.6 \kms, only every second channel is
  shown.\label{fig:chans}}
\end{figure}

\begin{figure}[t]
\epsscale{0.8}
\figurenum{2}
\figcaption{Continued --- As in previous panel. Contours in the
  bottom-right channel indicate (from the inside out) where the
  sensitivity has dropped to 90, 80 and 60 per cent of the central
  value. Note the different field centre compared to the previous
  panel. Also note the Galactic contamination in the $-11.2$ and
  $-8.0$ \kms\ channels.}
\end{figure}

\begin{figure*}[t!]
\plotone{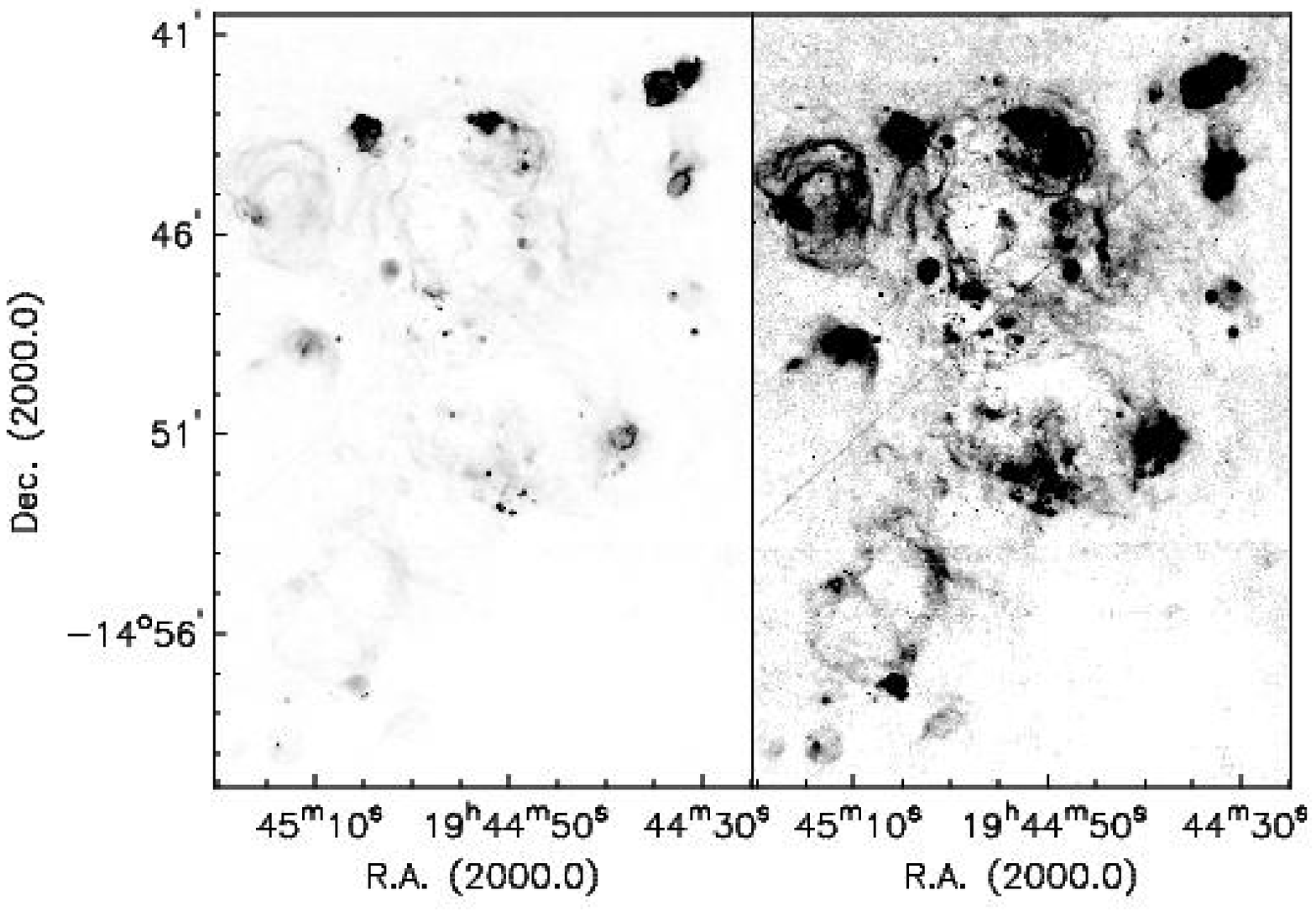}
\figcaption{The H$\alpha$ distribution in the central part of NGC6822
  as indicated in Fig.~\ref{fig:halfa}.  The left panel shows a
  low-contrast image of the H$\alpha$ distribution, clearly showing
  the population of bright H$\alpha$ regions, as well as the compact,
  high surface brightness regions. The right panel shows a
  high-contrast image of the same field emphasizing the network of
  low-surface brightness H$\alpha$ filaments found throughout the
  central part of NGC 6822.\label{fig:halfa_cen}}
\end{figure*}

\clearpage

\begin{figure*}[t!]
\figcaption{Outline of the SGS superimposed on various mass component
  distributions. The inner thick ellipse has a semi-major axis of
  250$''$.  The outer thick ellipse has a semi-major axis of
  430$''$. The thin dashed ellipses have semi-major axes of $170''$
  and $300''$. For description see text. Shown in the top-left panel
  is the distribution of \HI.  Top-right shows the H$\alpha$,
  bottom-left the distribution of blue stars, bottom right the $B-R$
  colour of the blue population (dark corresponds with a red colour,
  light with a blue colour).
\label{fig:hole2d}} \end{figure*}

\end{document}